\documentstyle[12pt]{article}
\newlength{\pubnumber} 

\catcode`\@=11
\@addtoreset{equation}{section}
\def\section{\@startsection{section}{1}{\z@}{3.5ex plus 1ex minus .2ex}
 {2.3ex plus .2ex}{\large\bf}}
\def\subsection{\@startsection{subsection}{2}{\z@}{2.3ex plus .2ex}
 {2.3ex plus .2ex}{\bf}}

\begin{document}

\begin{titlepage}
\samepage{
\setcounter{page}{1}
\rightline{IASSNS-HEP-97/68}
\rightline{CERN-TH/97-292}
\rightline{\tt hep-ph/9712322}
\rightline{December 1997}
\vfill
\begin{center}
 {\Large \bf Twenty Open Questions\\ in Supersymmetric Particle
Physics\footnote{
      To appear as an Overview Chapter in the review volume
        ``Perspectives on Supersymmetry'',
      edited by G. Kane, to be published by World Scientific.}
     \\}
\vfill
   {\large
    Keith R. Dienes$^{1,2}$\footnote{
     E-mail addresses: dienes@sns.ias.edu;  keith.dienes@cern.ch}
     $\,$and$\,$
      Christopher Kolda$^1$\footnote{
     E-mail address: kolda@ias.edu}
    \\}
\vspace{.16in}
 {\it  $^1$ School of Natural Sciences, Institute for Advanced Study\\
  Olden Lane, Princeton, N.J.~~08540~ USA \\}
\vspace{.05in}
  {\it $^2$ CERN Theory Division, CH-1211 Geneva 23, Switzerland \\}
\end{center}
\vfill
\begin{abstract}
  {\rm
   We give a brief overview of 20 open theoretical questions in supersymmetric
   particle physics.  The 20 questions we have chosen range from the GeV
   scale to the Planck scale, and include issues pertaining to the Minimal
   Supersymmetric Standard Model and its extensions,
   SUSY-breaking, cosmology, grand unified theories, and string theory.
   Throughout, our goal is to address those topics in which supersymmetry plays
   a fundamental role, and which are areas of active research in the field.
   This survey is written at an introductory level and is aimed
   at people who are not necessarily experts in the field.
   }
\end{abstract}
\vfill
\smallskip}
\end{titlepage}

\setcounter{footnote}{0}

\input epsf.tex

\def\Journal#1#2#3#4{{#1} {\bf #2}, #3 (#4)}
\def\NCA{\em Nuovo Cimento}
\def\NIM{\em Nucl. Instrum. Methods}
\def\NIMA{{\em Nucl. Instrum. Methods} A}
\def\NPB{{\em Nucl. Phys.} B}
\def\PLB{{\em Phys. Lett.}  B}
\def\PRL{\em Phys. Rev. Lett.}
\def\PRD{{\em Phys. Rev.} D}
\def\ZPC{{\em Z. Phys.} C}
\def\st{\scriptstyle}
\def\sst{\scriptscriptstyle}
\def\mco{\multicolumn}
\def\epp{\epsilon^{\prime}}
\def\vep{\varepsilon}
\def\ra{\rightarrow}
\def\ppg{\pi^+\pi^-\gamma}
\def\vp{{\bf p}}
\def\ko{K^0}
\def\kb{\bar{K^0}}
\def\al{\alpha}
\def\ab{\bar{\alpha}}
\def\be{\begin{equation}}
\def\ee{\end{equation}}
\def\bea{\begin{eqnarray}}
\def\eea{\end{eqnarray}}


\long\def\@caption#1[#2]#3{\par\addcontentsline{\csname
  ext@#1\endcsname}{#1}{\protect\numberline{\csname
  the#1\endcsname}{\ignorespaces #2}}\begingroup
    \small
    \@parboxrestore
    \@makecaption{\csname fnum@#1\endcsname}{\ignorespaces #3}\par
  \endgroup}
\catcode`@=12

\newcommand{\newc}{\newcommand}
\newc{\cpv}{CP-violating}
\newc{\bv}{{B\!\!\!\!/}}
\newc{\lv}{{L\!\!\!/}}
\newc{\order}{{\cal O}}
\newc{\beq}{\be}
\newc{\eeq}{\ee}
\newc{\gravitino}{{\widetilde{G}}}
\newc{\squark}{{\widetilde{q}}}
\newc{\Qsquark}{{\widetilde{Q}}}
\newc{\usquark}{{\widetilde{u}}}
\newc{\dsquark}{{\widetilde{d}}}
\newc{\csquark}{{\widetilde{c}}}
\newc{\tsquark}{{\widetilde{t}}}
\newc{\Lslepton}{{\widetilde{L}}}
\newc{\eslepton}{{\widetilde{e}}}
\newc{\smuon}{{\widetilde{\mu}}}
\newc{\selectron}{{\widetilde{e}}}
\newc{\stau}{{\widetilde{\tau}}}
\newc{\half}{\frac{1}{2}}
\newc{\kev}{\,\mbox{keV}}
\newc{\gev}{\,\mbox{GeV}}
\newc{\tev}{\,\mbox{TeV}}
\newc{\mev}{\,\mbox{MeV}}
\newc{\ev}{\,\mbox{eV}}
\newc{\gsim}{\lower.7ex\hbox{$\;\stackrel{\textstyle>}{\sim}\;$}}
\newc{\lsim}{\lower.7ex\hbox{$\;\stackrel{\textstyle<}{\sim}\;$}}
\newc{\mz}{m_Z}
\newc{\mpl}{M_{\rm Pl}}
\newc{\vev}[1]{\left\langle #1\right\rangle}
\newc{\vzero}{\vev{\Phi}_0}
\newc{\vth}{\vev{\Phi}_{T,H}}

\def\beq{\begin{equation}}
\def\eeq{\end{equation}}
\def\beqn{\begin{eqnarray}}
\def\eeqn{\end{eqnarray}}
\def\half{{\textstyle{1\over 2}}}
\def\NPB#1#2#3{{\it Nucl.\ Phys.}\/ {\bf B#1} (19#2) #3}
\def\PLB#1#2#3{{\it Phys.\ Lett.}\/ {\bf B#1} (19#2) #3}
\def\PRD#1#2#3{{\it Phys.\ Rev.}\/ {\bf D#1} (19#2) #3}
\def\PRL#1#2#3{{\it Phys.\ Rev.\ Lett.}\/ {\bf #1} (19#2) #3}
\def\IJMP#1#2#3{{\it Int.\ J.\ Mod.\ Phys.}\/ {\bf A#1} (19#2) #3}
\def\PRep#1#2#3{{\it Phys.\ Rep.}\/ {\bf #1} (19#2) #3}
\def\MSSM{{\rm MSSM}}
\def\beq{\begin{equation}}
\def\eeq{\end{equation}}
\def\beqn{\begin{eqnarray}}
\def\eeqn{\end{eqnarray}}
\def\KM{{affine}}
\def\rep#1{{\bf #1}}
\def\etal{{\it et al.}\/}
\def\inbar{\,\vrule height1.5ex width.4pt depth0pt}

\def\separator{{ \bigskip \bigskip \hrule \vskip 0.035 truein \hrule \bigskip
\bigskip }}
\def\IC{\relax\hbox{$\inbar\kern-.3em{\rm C}$}}
\def\IQ{\relax\hbox{$\inbar\kern-.3em{\rm Q}$}}
\def\IR{\relax{\rm I\kern-.18em R}}
 \font\cmss=cmss10 \font\cmsss=cmss10 at 7pt
\def\IZ{\relax\ifmmode\mathchoice
 {\hbox{\cmss Z\kern-.4em Z}}{\hbox{\cmss Z\kern-.4em Z}}
 {\lower.9pt\hbox{\cmsss Z\kern-.4em Z}}
 {\lower1.2pt\hbox{\cmsss Z\kern-.4em Z}}\else{\cmss Z\kern-.4em Z}\fi}

  \newc{\eg}{{\it e.g.}\/}
  \newc{\ie}{{\it i.e.}\/}
\hyphenation{su-per-sym-met-ric non-su-per-sym-met-ric}
\hyphenation{space-time-super-sym-met-ric}
\hyphenation{mod-u-lar mod-u-lar--in-var-i-ant}
\def\boxit#1{\vbox{\hrule\hbox{\vrule\kern3pt
\vbox{\kern3pt#1\kern3pt}\kern3pt\vrule}\hrule}}


\setcounter{footnote}{0}

At first glance, supersymmetry appears to be a theoretical
success story writ large.  With one simple
idea, such diverse issues as extending the Lorentz group, solving the
gauge hierarchy problem, coupling gauge theories to gravity, and
generating gauge coupling unification all seem
to fall into place. The lack of direct experimental evidence for supersymmetry
(SUSY)
dampens our enthusiasm somewhat, but only now (and over the next few years) are
experiments really beginning to probe the domain where SUSY should be expected
to be manifest.

Nonetheless, there is a price to be paid for the successes of SUSY,
and we mean more than simply the doubling of the Standard Model (SM)
particle spectrum. SUSY introduces into physics a host of new questions
which must be addressed, and hopefully, answered. Most of these questions were
once real problems --- problems which seemed to detract from, or
perhaps even invalidate,
SUSY as a viable fundamental symmetry of nature.
Although some of the questions presented here do not have attractive
solutions,
none of them, nor any others that we are aware of, threaten to rule out SUSY.
Instead, as competing answers to these questions have been found,
these questions become opportunities not only to simply discover SUSY, but
also to probe physics well beyond the scale of SUSY.

The questions that we will consider in this work have been chosen because
they satisfy a number of important criteria. First and foremost, each
defines an area of active research in the field; in many ways, the list of
questions
that follows forms a summary of current topics of interest in applying SUSY
to the physics of the SM.

Second, these are questions which are intrinsically supersymmetric and may
not even arise in the SM alone. Thus, generic questions in the
SM (such as the cosmological constant, inflation,
baryogenesis, and fermion mass hierarchies, to name a few)
are not included here. This is not to say that SUSY does not have
implications for these subjects, for it usually does, and when it does
it typically
recasts the terms of the debate completely by
redefining the spectrum of possible solutions.
However, in this article we will
concentrate on those issues which are intrinsically supersymmetric and whose
solutions one would hope to find in a complete description of a supersymmetric
SM. After all, it would be wonderful if SUSY could explain why the electron
is so much lighter than the top quark, but there is no obvious reason to
suppose that it will since the question itself is intrinsically
non-supersymmetric.

Third, these are questions whose answers may tell us a great deal about
physics beyond the MSSM, yet which might be probed using only
a combination of experimental measurements of the MSSM
and theoretical constraints. In this sense, these questions are
windows through which insight far beyond the weak scale
may be sought.

Because we focus primarily on questions that have direct applicability
to the physics of the SM or MSSM, a large number of interesting topics will
receive only abbreviated attention. For example, recent results on the
dynamics of SUSY gauge theories may have profound effects on how we
approach the MSSM in coming years (as they already have had on the question of
SUSY-breaking), but their current applications are limited.
Thus we will only touch upon this and related topics of a more formal
nature.

The open questions we have chosen to address are as follows.
First, at the lowest energy scales, we have selected a number of
open questions that pertain to the MSSM itself:
\begin{itemize}
\item {\sl Question \#1:}\/~~ Why doesn't the proton decay in $10^{-17}$
years?
  \item {\sl Question \#2:}\/~~ How is flavor-changing suppressed?
  \item {\sl Question \#3:}\/~~  Why isn't CP violation ubiquitous?
  \item {\sl Question \#4:}\/~~  Where does the $\mu$-term come from?
  \item {\sl Question \#5:}\/~~  Why does the MSSM conserve color and charge?
  \end{itemize}
  Next, we consider open questions pertaining to SUSY-breaking:
  \begin{itemize}
  \item {\sl Question \#6:}\/~~  How is SUSY broken?
  \item {\sl Question \#7:}\/~~  Once SUSY is broken, how do we find out?
  \end{itemize}
  Then, we consider two open questions pertaining to natural extensions of
  the MSSM:
  \begin{itemize}
  \item {\sl Question \#8:}\/~~  Can singlets and SUSY coexist?
  \item {\sl Question \#9:}\/~~  How do extra $U(1)$'s fit into SUSY?
  \end{itemize}
  Our next set of open questions addresses the interplay between
  supersymmetry and cosmological issues:
  \begin{itemize}
  \item  {\sl Question \#10:}\/~~  How does SUSY shed light on dark matter?
  \item  {\sl Question \#11:}\/~~  Are gravitinos dangerous to cosmology?
  \item  {\sl Question \#12:}\/~~  Are moduli cosmologically dangerous?
  \end{itemize}
  We then turn our attention to supersymmetric GUT's:
  \begin{itemize}
  \item  {\sl Question \#13:}\/~~  Does the MSSM unify into a SUSY GUT?
  \item  {\sl Question \#14:}\/~~  Proton decay again:  Why doesn't the
              proton decay in $10^{32}$
\phantom{{\sl Question \#14:}\/~~~$\,$} years?
  \item  {\sl Question \#15:}\/~~  Can SUSY GUT's explain the masses of
              fermions?
  \end{itemize}
  Next, we discuss some recent
  formal developments concerning SUSY and gauge theory:
  \begin{itemize}
  \item   {\sl Question \#16:}\/~~  N=1 SUSY duality:  How has SUSY changed
   our view of gauge
 \phantom{{\sl Question \#17:}\/~~$\,$} theory?
  \end{itemize}
  Finally, our last set of questions addresses supersymmetry at the very
  highest scales, in the context of string theory:
  \begin{itemize}
  \item  {\sl Question \#17:}\/~~  Why strings?
  \item  {\sl Question \#18:}\/~~   What roles does SUSY play in string
    theory?
  \item  {\sl Question \#19:}\/~~   How is SUSY broken in string theory?
  \item  {\sl Question \#20:}\/~~  Making ends meet:  How can we understand
  gauge coupling
\phantom{{\sl Question \#21:}\/~~~} unification from string theory?
  \end{itemize}

Finally, many of the topics that we shall discuss here will be
covered in much greater detail in the topical chapters of this book, and
we will try to indicate the relevant chapters as we go along.

\separator

\setcounter{footnote}{0}

  ~

\noindent
 {\large\bf Section I:~~ Open Questions in the MSSM}

 ~

The Standard Model forms the bedrock of modern high-energy physics,
and accurately describes all physical phenomena down to scales of $10^{-16}$
cm.
However, there are many possible ways of extending the SM down to smaller
length
scales.  These include extra gauge interactions, new matter,
new levels of compositeness (technicolor), and supersymmetry.
While supersymmetry does not succeed as an {\it explanation}\/ of the features
of the SM, it provides a remarkably robust extension to the SM which is in
agreement
with all experimental data.  This cannot be said for many other possible
extensions (such as, {\it e.g.}\/, the simplest versions of technicolor).
Moreover, it is quite possible and perhaps even likely
that other forms of potential new physics might appear at the
same energy scale as supersymmetry.
Thus, as a first step, it is important to investigate how
the structure of supersymmetry might be joined with that of the SM in a
cohesive framework.

Although there are various ways in which SUSY might be joined
with the SM, for simplicity one can pursue a {\it minimal}\/ construction,
and attempt to write down a Lagrangian which is the most {\it general}\/
 effective Lagrangian for the {\it minimal}\/ extension of the
SM which is invariant under SUSY transformations up to {\it soft-breaking}\/
terms.
This then results in
the Lagrangian of the Minimal Supersymmetric Standard Model (MSSM).
We will not attempt to fully define the MSSM, its field content, or its
Lagrangian,
but instead we refer the reader to any of several standard
references~\cite{mssm},
or to the article of S.~Martin in this volume.
We will therefore say only a few
words of a general nature.

The minimal extension of the SM with unbroken SUSY is a simple model with
fewer free parameters than the SM itself, despite the large number of new
fields. It is only in the breaking of SUSY that the number of free parameters
becomes large --- but of course, it is only in the breaking of SUSY that
the model can even attempt to describe nature as we observe it. In the SM,
the field content along with the gauge symmetries serve to provide a number
of accidental symmetries at the renormalizable
level.  These include, for example, baryon number $B$, and lepton number $L$.
These symmetries also serve
to forbid flavor-changing neutral currents (FCNC's) up to small corrections
arising from Yukawa couplings in loops. The MSSM shares neither of these
properties. As we will discuss, the most general MSSM would have the proton
decay
with a weak-interaction lifetime and large FCNC's. The reason is this same
proliferation of fields and free parameters. Even after imposing symmetries
to forbid the fast proton decay (see Question \#1 below), one finds~\cite{ds}
that the MSSM contains
106 new, independent (real) parameters above and beyond those of the
SM.  These consist of
26 masses (resulting from 12 squark, 9 slepton, and 3 gaugino
masses, plus $\mu$- and $B_\mu$-terms), 37 mixing angles, and 43
CP-violating phases. Understanding, constraining, and ultimately measuring
these parameters is one of the primary goals of the SUSY program in particle
physics.

Of course, the MSSM is unlikely to be the end of the story.
Therefore, although we will begin this article by considering open questions
within the MSSM, we will later allow this structure to expand
by considering new singlets, extra gauge symmetries, grand unification,
and ultimately embeddings within string theory.

One recurring feature in many of the open questions in this article
is the question of ``naturalness'': why is some coupling (or mass) very small
or
even zero when it need not have been zero {\it a priori}\/?
More precisely, this is a question of ``Dirac naturalness.'' The idea of
Dirac naturalness is built on
the supposition that in any physical system, all couplings and
interactions which are not otherwise forbidden should be allowed, and
that all ratios of couplings, as well as all ratios of masses, should be
${\cal O}(1)$. We generally find that a theory is Dirac natural if some
exact symmetry exists which forbids the undesired couplings.
Note that Dirac naturalness is not to be confused with ``'t~Hooft
naturalness'' or ``technical naturalness,'' which is the problem besetting
the Higgs sector of the Standard Model. In the latter case
one seeks to understand small
numbers or ratios, such as the ratio of the weak to Planck scale, in terms
of approximate symmetries. The role of radiative corrections is very important
for 't~Hooft naturalness, because even if one could choose the ratio of two
couplings or two masses to be far from unity at tree level, without some
approximate symmetry at play there would be no reason for this ratio to
persist beyond tree level. The chiral symmetry of the SM fermions is a
classic example of this phenomenon:
an approximate symmetry protects the fermion masses from
receiving corrections proportional to heavy mass scales.

Without SUSY, there is nothing like a chiral symmetry to protect scalar
masses from heavy mass scales.
But with SUSY, the chiral symmetry in the fermionic
sector protects the scalars too. This is a general feature of SUSY ---
couplings and masses which are SUSY-preserving are automatically natural
 {\it \`a la}\/ 't~Hooft
even if they are unnatural {\it \`a la}\/ Dirac. Thus
when we speak of naturalness in the context of SUSY, we will generally be
referring to Dirac naturalness, for which SUSY provides no automatic
solutions.

\section*{\fbox{Question \#1}~~ Why
doesn't the proton decay in $10^{-17}$ years?}

As discussed in the article of S.~Martin in this volume,
the most general, renormalizable superpotential for the MSSM can be organized
into three pieces, $W=W_0+W_\bv+W_\lv$, where
\bea
W_0&=&y^U_{ij}Q^iu^jH_U+y^D_{ij}Q^id^jH_D+y^E_{ij}L^ie^jH_D+\mu H_UH_D \\
W_\lv&=&\lambda_{ijk}Q^id^jL^k+\lambda'_{ijk}e^iL^jL^k+\mu'_iH_UL^i \\
W_\bv&=&\lambda''_{ijk}u^id^jd^k~.
\eea
The first piece preserves the global $B$ and $L$ quantum numbers of the SM,
while the other two each violate one of either $B$ or $L$. This is to be
contrasted with the case of the SM in which one can obtain
$B$-violating operators
only by going to dimension six, or $L$-violating operators only by going to
dimension five. Allowing
simultaneous $B$- and $L$-violation would be a disaster. For example, given
non-zero
$\lambda$ and $\lambda''$, one can form a four-fermion
operator $QudL$ which can mediate
proton decay and is only suppressed by squark masses~\cite{dg,pdecay},
not $M_{\rm GUT}$ or $\mpl$.

There is a simple remedy for this:  one
can set $W_\bv=W_\lv=0$ by hand.
In a non-supersymmetric theory, this would be 't~Hooft unnatural unless
there existed some exact global or gauge symmetry to ensure
that these couplings remain zero.
In a supersymmetric theory, however, couplings in the superpotential are always
't~Hooft natural due to the ``non-renormalization'' theorem, which says that
any couplings in $W$ that are set to zero will remain zero to orders in
perturbation theory.
Nonetheless, setting otherwise-allowed couplings to zero is always Dirac
unnatural and so we might look for a symmetry that arises when these
couplings vanish. The obvious candidates,
global $B$ and $L$ symmetries, are probably not suitable to play this role
since we expect from a number of arguments (\eg, involving GUT's, $\nu$
masses, cosmological baryon asymmetry)
that they will be broken by non-renormalizable or non-perturbative terms.

Fortunately it is easy to invent a new discrete symmetry which simultaneously
forbids all the unwanted terms and which allows those that are
phenomenologically necessary. This is ``matter parity,'' a $\IZ_2$ symmetry
under which all the matter fields
$(L,e,Q,u,d)$ are odd and the Higgs fields are even~\cite{drw}. An added
benefit of such a symmetry is the exact stability of the lightest
supersymmetric particle (LSP),
even against higher-order interactions, thereby providing
a candidate for the (cold) dark matter in the universe.

Though matter parity may provide a ``Dirac natural'' solution to the proton
decay problem,
there is a more recent definition of naturalness
which matter parity does not necessarily satisfy.
We shall refer to
this new definition, which
comes out of ideas in
string theory and quantum gravity,
as ``local naturalness.'' Whereas Dirac naturalness would allow any exact
global symmetry to forbid the unwanted couplings,
local naturalness requires that
these symmetries be {\it gauge}\/ symmetries, or at least the indirect
consequence of
an underlying gauge symmetry~\cite{wormhole}.
This requirement follows from the result/belief
that in a theory of quantum gravity, there may be effects  (potentially
arising from Planck-scale physics)
which violate any and all symmetries of the theory which are not gauged or
somehow protected by a gauge symmetry. These protected symmetries include
global and discrete subgroups of gauge symmetries. And though these
Planck-scale
effects might be small, suppressed by powers of $\mpl$, in some cases this
may be enough to violate known constraints. In the case of proton decay, such
a violation occurs. It would be desirable, then, to
embed matter parity into a gauge symmetry.

In order to see how this might be done, note that the matter parity $P_M$
of any MSSM field with baryon and lepton numbers $B$ and $L$ respectively
can be written as
\beq
P_M=(-1)^{3(B-L)}.
\eeq
Matter parity can therefore easily be
accommodated as a discrete subgroup of a
gauged $U(1)_{B-L}$ symmetry, something which appears in many extensions
of the SM. In fact, the action of the full $U(1)_{B-L}$ symmetry would  be to
forbid
exactly the unwanted terms in $W$. If this selection rule is to survive
the breakdown of the $U(1)_{B-L}$ symmetry, then one need only require that all
order parameters (\eg, Higgs VEV's) carry {\it even}\/ integer values of
$3(B-L)$.
This restriction can also be generalized to groups containing $U(1)_{B-L}$;
for example, in $SO(10)$ one finds~\cite{martin} that matter parity occurs as a
discrete
gauge symmetry after GUT-breaking if that breaking is done without giving
VEV's to spinor representations
(\eg, ${\bf 16, 144, 560}\ldots$).

The question of how one
can embed such a discrete symmetry into a gauge symmetry can be further
generalized by asking whether or not the set of discrete charges in the MSSM
is ``anomaly-free,'' {\it i.e.}\/, whether or not these charges
obey certain constraints which
allow them to be interpreted as arising from a
non-anomalous gauge symmetry (as in the
example above). One can in fact show that matter parity is the only
(generation-independent) $\IZ_2$ symmetry which is anomaly-free and prevents
dimension-four proton decay~\cite{ir}.

Of course, the original
argument against allowing non-zero $W_\bv$ and $W_\lv$ was
that both could not be allowed without leading to rapid nucleon decay. This
argument suffices to forbid one or the other, but not both. In fact,
much work has been done considering extensions of the MSSM with so-called
``$R$-parity violation'' in which either $W_\bv$ {\it or}\/ $W_\lv$ is
non-zero,
but not both. The price one pays is that the LSP is no longer stable and
so there is no easy candidate for the dark matter. However, the phenomenology
and motivations of $R$-parity violation are too complicated to discuss here;
for details, see the contribution of H.~Dreiner to this volume.

Thus far, we have said nothing about proton decay from higher-dimension
operators. In particular there are many terms of dimension five
 {\it which are invariant under matter parity}\/ which can mediate proton
decay.
For example, let us consider the superpotential term
$W=(\frac{\eta}{M})QQQL$.
Even if one identifies $M$ with the Planck mass, current bounds on the
proton lifetime require that $\eta$ be unnaturally small:
$\eta\lsim10^{-7}$. One could consider alternative $\IZ_N$ symmetries
for the superpotential, and in fact one particular alternative~\cite{ir}
stands out:
a $\IZ_3$ symmetry called ``baryon parity.'' Under baryon parity the MSSM
fields $(\Qsquark,\usquark,\dsquark,\Lslepton,\eslepton,H_U,H_D)$ have
$\IZ_3$ charges (0,2,1,2,2,1,2) respectively. Baryon parity has many
interesting properties: it is the only (generation-independent)
$\IZ_3$ symmetry of the MSSM without discrete gauge anomalies; it prevents
dimension-four proton decay by forbidding $W_\bv$; it allows the $\mu$-term;
it allows $W_\lv$ and therefore neutrino masses; it forbids dimension-five
proton decay; but it also allows the decay of the LSP.

Is matter parity, or any of its competitors, a real symmetry of nature?
This is essentially an experimental question, but its implications go far
beyond questions of detection signals, for these symmetries teach
us about dark matter and the ultimate fate of the universe, and
also provide insights into the symmetry structure of the MSSM at very high
energies.

\section*{\fbox{Question \#2}~~ How is flavor-changing suppressed?}

There is no reason to expect that the mass and interaction eigenstates
of the SM fermions coincide; that is, the quark and lepton mass matrices
need not be diagonal in the interaction eigenbasis.
In fact we know that they are not diagonal (since $\theta_c\neq0$).
One obviously expects the same to be true of the scalars of the MSSM. But
before SUSY-breaking, one can expect that at least the mass matrices of
fermions and their scalar superpartners will be diagonal in the same basis.
This need not be true after SUSY-breaking.

In the interaction basis, non-diagonal mass matrices would seem
to lead to large flavor-changing neutral currents (FCNC's).
Within the SM, it is the GIM mechanism which prevents this from occurring.
Flavor-independent neutral current (NC) gauge interactions couple gauge bosons
to
propagating fermions
and their conjugates, each rotated from their interaction basis by matrices
$U$ and $U^\dagger$ respectively. As long as $U$ is unitary, then $U^\dagger
U={\bf1}$ and the NC gauge interaction conserves flavor. The same holds for
the scalar partners which are rotated by unitary $\widetilde{U}$ and
$\widetilde{U}^\dagger$. Thus
gauge bosons do not induce FCNC's for particles or their superpartners. The
Problem arises with gauginos, which couple to both a particle {\it and}\/ its
superpartner simultaneously. In this case the coupling has the form
$\widetilde{U}^\dagger U$, which need not be diagonal. Thus gauginos can
generate FCNC's~\cite{dg,fcnc}. (This is not a complete disaster because
for flavor-changing processes involving only fermions on external lines,
gaugino contribution can arise only beyond the tree-level.)

Because FCNC's are suppressed by GIM in the SM, the dominant source for FCNC's
in low-energy processes may come from SUSY.
The tightest bounds on FCNC's presently
come from $K^0-\overline{K^0}$ mixing. Requiring that the MSSM prediction
for the $K^0-\overline{K^0}$ mass difference not exceed the measured value
yields limits of the form~\cite{ggms}:
\beq
\frac{{\cal A}^2}{m^2_\Qsquark}\left(\frac{\delta m^2_\Qsquark}{m^2_\Qsquark}
\right)^2~\leq~5\times10^{-9}\,\gev^{-2}
\eeq
where ${\cal A}^2$ is a product of angles which rotate from the quark mass
basis to the squark mass basis, and $\delta m^2_\Qsquark$ is the mass
difference between the $\widetilde{d_L}$ and $\widetilde{s_L}$ squarks.
There are corresponding limits on $\widetilde{d_R}-\widetilde{s_R}$ splittings
as well as mixed left-right limits.

Given the above constraint, it is clear that
if the mass splittings are $\order(1)$ and the angles take average
values (${\cal A}^2\sim1/20$), the squark mass scale must be $\gsim3\tev$.
Similar bounds exist from $D^0-\overline{D^0}$ and $B^0-\overline{B^0}$
mixing, and
in the slepton sector from processes such as
$\mu\to e\gamma$. However
it is worth noting that bounds on FCNC's tend to constrain
only the first two generations of squarks and sleptons; the third
generation is rather weakly constrained at present.

There are three primary proposals for solving the SUSY flavor problem:
degeneracy, alignment, and decoupling.

Degeneracy attempts to solve the flavor problem by positing that squarks
and sleptons of a given flavor are mass-degenerate~\cite{dnw,ggms},
\ie, $\delta m^2_\Qsquark=0$.
(One way to see that this would forbid FCNC's
is to note that if, for example, the $\widetilde{d}$ and $\widetilde{s}$
were to have equal mass, they would exactly cancel each other's contributions
to any flavor-changing process.) This extension of the GIM mechanism to the
SUSY sector is often referred to as ``super-GIM'':
$m^2_\usquark=m^2_\csquark\simeq
m^2_\tsquark$ for L,R squarks separately, likewise $m^2_\selectron=
m^2_\smuon\simeq m^2_\stau$.
Given the weaker bounds on FCNC's involving the third generation,
the final equalities need only be approximate.

Degeneracy models come in a variety of flavors themselves (for more details
see Question \#7). For many years,
models based on supergravity (SUGRA) as a mediator of SUSY-breaking were
considered members of this family. Now it is understood that there exist
a number of rather generic phenomena which break degeneracy in SUGRA models,
including non-minimal K\"ahler potentials, GUT effects, and superstring
thresholds. Even if all of these could be ruled out, it is important to
realize that {\it sources of flavor physics between the Planck and weak
scales tend to violate the degeneracy.} This is a generic feature: if flavor
physics occurs below the scale at which SUSY-breaking is communicated to the
SM, scalar mass degeneracies will tend to be spoiled.
An alternative to SUGRA mediation is
gauge-mediated SUSY-breaking (GMSB). In GMSB two important things happen:
the scalars of the MSSM receive their soft masses from SM gauge interactions,
which are by definition flavor-universal, and these masses are communicated at
scales often very close to the weak scale so that there is little room for new
physics to spoil the degeneracy. Finally
it has also been suggested that the flavor physics itself arises from
non-abelian global symmetries (``horizontal'' symmetries)
under which the families of the SM/MSSM form
non-trivial representations. These models usually predict some combination
of degeneracy among the flavors and alignment, the next mechanism. One perhaps
noteworthy aspect of non-abelian horizontal symmetries is that they can
be gauged only in special cases~\cite{bb},
because the broken Cartan generators of a local symmetry typically
generate $D$-terms which destroy the mass degeneracy that one worked so
hard to obtain in the first place.

Alignment solves the flavor problem not by setting $\delta\widetilde{m}^2=0$
as with degeneracy, but by enforcing $U=\widetilde{U}$ (or equivalently
${\cal A}=0$) to a very high accuracy~\cite{ns}.
Here the flavor physics is typically generated by one or more
$U(1)$ gauge interactions. Models of abelian horizontal symmetries seek to
tie the generation of the scalar soft masses to that of the fermion
mass/Yukawa matrices. All of these models generate the hierarchies in the
fermion masses as powers of a small expansion parameter,
usually the ratio between some
flavor-violating VEV's and the UV scale of the theory. Particularly interesting
among these models are those in which the flavor $U(1)$ is
pseudo-anomalous~\cite{nw}
--- the anomalies in its fermionic sector are cancelled by non-linear
transformations of the dilaton/axion superfield as in the Green-Schwarz
mechanism of string theory. Such a $U(1)$ must be broken just below the
string scale via a one-loop induced Fayet-Iliopoulos term, and it is the ratio
of this breaking scale to the string scale that provides the expansion
parameter for building the mass hierarchies.

Finally, decoupling has been proposed as a solution to the flavor
problem~\cite{ckn}.
Here one simply makes the first two generations of sparticles
heavy enough (typically $10-100\tev$) so that their contributions
to FCNC processes vanish.
But what about the 't~Hooft naturalness problem in
the SM Higgs sector that SUSY can
solve only if the superpartners are light ($\lsim1\tev$)?
Recall that the troublesome diagrams involving scalars were
all suppressed by Yukawa couplings. If we make the reasonable assumption that
naturalness requires the gauginos, higgsinos, {\it and the third generation
of squarks and sleptons}\/ to lie below $1\tev$, then all the other squarks
and sleptons can have masses as large as $(m_t/m_{q,\ell})\times(1\tev)$
without violating naturalness constraints. (Here $m_{q,\ell}$ is the
appropriate partner quark/lepton mass.) Once again
one takes advantage of the fact that the bounds on third generation FCNC's
are weak in order
to allow large mass splittings between the third generation and
the first two. One should note, however, that there exist problems with the
behavior
of the decoupling scenario at two-loop order~\cite{ahm};
it is not clear at present that such a scenario can be made to work without
inadvertently breaking QCD and/or QED.

Once SUSY is discovered, it may not take long to discern which of these
paths nature has chosen to follow. Observation of scalar partners should
quickly tell us whether they are degenerate or not (\ie,
super-GIM or aligned); likewise, if only the third generation is found, we
learn
that the other spartners must be
very heavy (\ie, decoupled). However, it will take
much more experimental and theoretical work to determine just how each
of these choices is concretely realized.

\section*{\fbox{Question \#3}~~ Why isn't CP violation ubiquitous?}

In the MSSM there are 43 physical \cpv\ phases above and
beyond that of
the SM. (In this discussion, we will ignore $\theta_{\rm QCD}$.)
Unlike the single SM phase, which does not typically engender large
\cpv\ effects\footnote{Even in the SM
  it is not always true that large \cpv\ observables are lacking; for example,
  in $B-\overline{B}$ mixing, \cpv\ effects can be $\order$(1) since $J$
appears
  divided by small quark mixing angles.}
because in physical processes it always comes in proportional
to the small Jarlskog parameter $J$,
the phases of the MSSM can show up in
large, easily observed, and easily constrained experimental processes.
The tightest constraints come from \cpv\ in the kaon system and
the electric dipole moment of the neutron. The former receives SUSY
contributions only if there are also flavor-changing SUSY contributions such
as those discussed above; the latter exists even if SUSY is flavor-preserving.
(For a review, see Ref.~\cite{gnr} and the contribution of A.~Masiero and
L.~Silvestrini to this volume.)

Let us consider the second case first, with all SUSY flavor-changing effects
put to zero through universal soft masses. For simplicity let us make the
usual assumption that
the trilinear couplings are all proportional to the Yukawas, and that
the gaugino masses are universal at some scale. Then there remain only two
physical phases beyond the SM associated
with some combination of the $\mu$-term, the $B_\mu$-term, the $A$-terms, and
the gaugino masses:
\beq
\phi_A=\arg(A^*M_3)\quad\quad\quad\quad\phi_B=\arg(B_\mu^*\mu M_3).
\eeq

At one-loop order, gluinos and squarks
can contribute to the electric dipole moment (EDM) of quarks, and then in turn
nuclei. The EDM of the neutron can be calculated to be~\cite{fpt} (for $M_3
\simeq
m_\squark\equiv\widetilde{m}$):
\beq
d_N\simeq 2\left(\frac{100\gev}{\widetilde{m}}\right)^2\sin(\phi_A-\phi_B)
\times 10^{-23}\,e\,{\rm cm},
\eeq
where experimentally $d_N<1.1\times10^{-25}\,e\,$cm. Clearly,
if the phases $\phi_{A,B}$ take values of $\order(1)$, then the sparticles
must be heavier than 1\tev; or for sparticles around $100\gev$, the
phases must be $\lsim10^{-2}$. In either case (heavy sparticles or small
phases) some degree of unnaturalness is introduced.

In the most general case where flavor violations are also allowed, not only
does the number of physical \cpv\ phases increase, but the bounds
from observables become much stronger.  For example, gluinos and squarks
can appear in the internal lines of box diagrams contributing to
$\epsilon_K$ (\ie,
the imaginary part of the $K^0-\overline{K^0}$ mixing amplitude).
One finds for $ M_3\simeq m_\Qsquark\simeq m_\dsquark\equiv\widetilde{m}$
that~\cite{ggms}
\beq
 \epsilon_K\simeq \left(\frac{10^{10}\gev^2}
 {\widetilde{m}^2}\right)
  {\rm Im}\,  \left(\frac{\delta m^2_\Qsquark}
    {m^2_\Qsquark}\,\frac{\delta m^2_\dsquark}{m^2_\dsquark}\right)~.
\eeq
Experimentally, $\epsilon_K$ is known to be about $2.3\times10^{-3}$.
Thus, in order to allow $\order(1)$ mass splittings and phases,
the squarks and sleptons must have masses $\gsim1000\tev$! Conversely, in order
to allow masses below $1\tev$, either the mass splittings or the \cpv\
angles must be made unnaturally small.

The CP problem is solved in ways that are similar to those that solve the
flavor
problem. For the $\epsilon_K$ problem, degeneracy appears to be an attractive
solution (decoupling from $\epsilon_K$ would seem to put
too heavy a burden on any theory, pushing squarks to $1000\tev$). The
EDM problem is not so simply solved, but then it is not quite as
serious. Decoupling {\it can}\/ eliminate this problem; it is also solved
in some classes
of very minimal gauge-mediated supersymmetry-breaking (GMSB)
models~\cite{CPinGMSB}. One particularly interesting
line of inquiry has involved attempts to use this CP-violation as the source
needed during baryogenesis~\cite{cw};
in this case one obtains strong constraints on the
parameters on the MSSM and thus a highly predictive model that will soon
be tested.

\section*{\fbox{Question \#4}~~ Where does the $\mu$-term come from?}

The $\mu$-term of the MSSM prompts another question of Dirac naturalness: how
do we induce a parameter to take a value far below its ``natural'' scale?
For the $\mu$-parameter of the MSSM,
\beq
W=\cdots+\mu H_UH_D,
\eeq
the natural value is the UV cutoff of
the MSSM. This is not a statement about radiative corrections {\it \`a la}\/ 't
Hooft,
for in SUSY models the non-renormalization theorem will protect a weak-scale
$\mu$-parameter from any large corrections. Rather, it is a question of why
a mass parameter which is SUSY-invariant and $SU(3)\times SU(2)\times
U(1)$-invariant would have a value typical of SUSY-breaking, or SM-breaking,
masses.  {\it A priori}\/, we would expect it to have a value of order
the scale at which $H_UH_D$ no longer forms a gauge singlet, or the Planck
scale, whichever is smaller. Within the context of a GUT model, the
problem is exacerbated. In $SU(5)$ parlance, the $\mu$-term provides the
mass for the complete $\bf5$ and
$\bf\overline5$ of Higgs, thereby becoming entangled in the famous
doublet-triplet splitting problem.

Phenomenologically, we know that $\mu\sim\mz$ because minimization of the MSSM
Higgs potential yields the result
\beq
\mu^2=\frac{m^2_{H_D}-m^2_{H_U}\tan^2\beta}{\tan^2\beta-1}-\half\mz^2
\eeq
where all the masses on the right side are $\sim\mz$. Only a
gross fine-tuning would allow values of $\mu$ very different than $\mz$.
We could in principle set $\mu\equiv 0$ by invoking a Peccei-Quinn symmetry,
but the
by-product of this would be a standard axion, already ruled out by direct
searches.

There is an almost-default solution to the $\mu$-problem which goes
by the acronym NMSSM (Next-to-Minimal$\ldots$). In this model, a gauge singlet
$N$ is introduced whose role is to produce a $\mu$-term through its VEV.
The superpotential would have the form:
\beq
W=\lambda NH_UH_D+\lambda' N^3
\eeq
while $N$ itself is presumed to have a soft (mass)${}^2$ term which is
negative. (Such a negative (mass)${}^2$ can actually arise naturally if
$\lambda$ or $\lambda'$ is large,
since it will drive the soft mass term negative in the
infrared, just as occurs for the $H_U$ mass term in
radiative electroweak symmetry breaking.) Obviously a $\mu$-term arises:
$\mu=\lambda\vev{N}$. Several terms contribute to a $B_\mu$-term, including
the trilinear soft term $\lambda A_N\vev{N}H_UH_D$, and $F_N$ via $B_\mu\sim
F_NH_UH_D\sim\lambda'\vev{N^2}H_UH_D$.
However, the singlet solution to the $\mu$-problem is not without problems,
as we will discuss further after Question \#8.

Within SUGRA there is actually a more attractive solution, known as the
Giudice-Masiero mechanism~\cite{gm}. If the $\mu$-term is forbidden by some
symmetry (say a discrete symmetry) which is violated in the hidden sector,
then a $\mu$-term can arise through a
non-minimal K\"ahler potential: $K=\cdots\,+(S/\mpl)H_UH_D$, where the ellipsis
represents the canonical terms and $S$ is a hidden sector field with
$F$-term $\langle F_S\rangle=\mz\mpl$. Then
\beq
\int d^4\theta \frac{S}{\mpl}H_UH_D=
\int d^2\theta \frac{F_S}{\mpl}H_UH_D \equiv
\int d^2\theta\, \mu H_UH_D.
\eeq
In this way the $\mu$-term, which is SUSY-preserving, is actually
tied to the breaking of SUSY and thus naturally $\sim\mz$.
The corresponding value of $B_\mu$ will then also be $\sim\mz^2$.

Within non-SUGRA models, there are no such simple solutions. In particular,
GMSB models struggle with a severe $\mu$-problem. In generic models
one typically finds $B_\mu/\mu^2\gg1$ where one needs $B_\mu\sim\mu^2$
phenomenologically. In some special cases one finds $B_\mu/\mu^2\ll1$.
Here one would expect an axion, but one-loop corrections to $B_\mu$
pull the axion mass above $\mz$. The hallmark of such a
scenario~\cite{CPinGMSB}
is a very large value for $\tan\beta$ $(\sim50)$.

\section*{\fbox{Question \#5}~~ Why does the MSSM conserve color and
 \phantom{\fbox{Question \#5}~~~~$\,$} charge?}

In the SM, the only field which can receive a VEV
is the Higgs field, and because of its quantum numbers, a Higgs VEV uniquely
breaks $SU(3)\times SU(2)\times U(1)\to SU(3)\times U(1)$, thereby
preserving QCD and
QED. In the MSSM there are a large number of charged and colored scalars in
addition to the Higgs, any of which could
receive a VEV and break the gauge group even further. Whether or not
this occurs is a function of the potential felt by these scalars.
Unfortunately,
minimizing this potential can be an arduous task, complicated by the
large numbers of scalar fields. (For a full discussion of efforts on this
front, see Ref.~\cite{casas}
and the contribution of A.~Casas to this volume.)

Scalar potentials in SUSY receive contributions from three sources:
$D$-terms, $F$-terms, and soft-breaking terms. The first of these provides
the quartic terms $V\sim\lambda\varphi^4$ with $\lambda\geq0$. For fields
which carry some gauge charge, $\lambda=0$ can occur along only special
directions in field space known as ``$D$-flat directions'' or ``$D$-moduli.''
Along
a flat direction the quartic potential takes the form $V\sim(\varphi_1^2-
\varphi_2^2)^2\to0$ as $|\varphi_1|\to|\varphi_2|$, and so
the potential far from the origin may not be well-behaved (\ie,
$\varphi$ may run off to infinity).
Whether or not this occurs will depend on the $F$-terms and soft terms.

The $F$-terms in turn contribute quadratic, cubic, and quartic terms to the
potential.
Because these terms
are supersymmetric, the portion of the potential due to $F$-terms
is positive semi-definite. This does not mean, however, that the minimum
of the potential lies at the origin. Rather, it means
that directions in field
space with non-zero quartic $F$ contributions will be well-behaved far away
from the origin. Still, there will be a subset of the $D$-flat directions
which are also $F$-flat and whose behavior will be completely controlled
by soft-breaking terms.

But the soft mass contributions are problematic, as they can have either
sign. Because they contribute to only the quadratic and cubic pieces of
$V$, one can analyze their structure most readily along flat directions
in which the quartic pieces all vanish. Then one finds two distinct types
of problems which may arise: potentials which are charge- and color-breaking
(CCB) at their minima, or potentials which are unbounded from below (UFB).

CCB most readily occurs along directions which are $D$-flat, though not
necessarily
$F$-flat; then the $\varphi^4$ contributions to the potential are suppressed
by Yukawas.
The canonical example~\cite{ccb}
of CCB involves only the fields $H_U$, $\Qsquark$ and
$\usquark$ which have a $D$-flat direction in which $|H_U|=|\Qsquark_u|=
|\usquark|$. The potential along this direction receives dangerous cubic
contributions from the soft trilinear terms $\lambda_u A_u\Qsquark H_U
\usquark$ which can dominate over the small residual quartic terms
(proportional to Yukawa couplings since this is not an $F$-flat direction)
out to large field values. The condition
that a secondary (and deeper) minimum not be generated away from the origin
then results in the famous bound:
\beq
|A_u|^2\leq3(m^2_\Qsquark+m^2_\usquark+m^2_{H_U}+\mu^2).
\eeq
In principle, bounds such as this can be derived along every $D$-flat direction
of the MSSM. It is clear that rigorously analyzing such a possibility is
hopeless
when one considers that the space of all $D$-flat directions is itself
37-complex dimensional!

The appearance of UFB directions is also common, occurring along directions
which are both $D$- and $F$-flat.
The usual example
is found right in the Higgs potential along the direction $|H_U|=|H_D|$.
As with all UFB potentials, there is no quartic
contribution to the potential along this direction
(nor in this example is there a cubic piece). Stability of the potential then
requires that the quadratic pieces be positive semi-definite, \ie,
\beq
m^2_{H_U}+m^2_{H_D}+2\mu^2\geq 2|B_\mu|.
\eeq
Once again, bounds such as these are fairly generic --- the space of directions
which are both $D$- and $F$-flat in the MSSM is $29$-complex
dimensional~\cite{gkm}.

Can these considerations prove useful beyond providing bounds on soft
parameters? The answer to this question depends somewhat on the values of
the soft masses which are measured experimentally. One can turn around the
above analysis and ask: What would it mean if the measured masses violated
a CCB/UFB constraint? In the case of UFB directions, the stability of the
potential must be rescued, either by non-renormalizable operators coming
from new physics
(\eg, $V\sim\varphi^6/M^2$) or by one-loop contributions to the effective
potential (\eg, $V\sim\mbox{STr}\,m^4(\varphi)\log[m^2(\varphi)/Q^2]/64\pi^2$).
In either case, the result is generally a bounded potential with CCB
VEV's well above the weak scale.

However, the existence of a global
CCB minimum below that of the SM does not necessarily imply that we should
be in it. It is
entirely possible that, at the end of inflation and reheating,
the universe found itself
in the current vacuum even though there is a deeper vacuum elsewhere. If the
barrier between the two vacua is high, the time scale for our universe
to tunnel to the new vacuum may be much larger than its current age.
But why would the universe end up in the ``wrong'' vacuum to begin with?
Presumably the answer lies in how the CCB minima are lifted by
finite-temperature effects relative to the SM-like minimum~\cite{kls}.
It may also depend on whether or
not some late (weak-scale) inflation occurred and what its reheating
temperature was. Once the universe ended up in the SM-like vacuum,
transitions to the CCB vacuum would be exponentially suppressed by the
height of the intervening barrier, easily leading to a metastable (but very
long-lived) universe.
Thus the appropriate question raised by discovering violation of the CCB bounds
may well be cosmological rather than directly experimental.

\separator

\setcounter{footnote}{0}

 ~

\noindent
{\large\bf Section II:~~ Open Questions on SUSY-Breaking}

 ~

All of the questions we have discussed thus far
have begun with the structure of the MSSM.
Implicit in that construction is the fact that
as a symmetry of nature,
SUSY
makes some profoundly (and obviously) wrong
predictions. In supersymmetric theories, there is
an absolute correspondence between fermions and bosons that is not manifest
experimentally. Specifically, SUSY requires a spectrum
in which every fermionic degree of freedom has a bosonic
counterpart with identical mass and quantum numbers.
Therefore, in order for SUSY to play any role in low-energy physics, it must
clearly
be broken (or hidden), just as is the $SU(2)$ gauge symmetry of weak
interactions.

\section*{\fbox{Question \#6}~~  How is SUSY broken?}

Once it is clear that SUSY must be broken, the next logical step would
seem to be to find some way in which to break SUSY spontaneously.
Why spontaneously? Our experience
with gauge interactions teaches us that Ward identities, and with them all
of the desirable properties of symmetries, are preserved after the
symmetry is broken {\it only if}\/ the symmetry-breaking is done
spontaneously. For gauge
symmetries, this means that renormalizability and unitarity are lost for
explicit breaking; for SUSY, it is the cancellation of the quadratic
divergences that would be lost. There is also another, more philosophical,
reason for demanding spontaneous SUSY-breaking:
symmetries which are explicitly broken are not symmetries at all
(even if they may be useful for classification purposes),
while symmetries which are
broken spontaneously are still symmetries of the theory --- they are just
not symmetries of the vacuum state of the theory.

It is clear that breaking SUSY spontaneously entails
adding to the SM some fields and their interactions to act as
a SUSY-breaking sector, just as the Higgs and its potential are added to the
SM. Can the fields of the MSSM play this role themselves? For a number
of reasons, it turns out that they cannot.
 First, of the MSSM fields that could receive a SUSY-breaking VEV,
it is only
the Higgs and sneutrinos which can do so without breaking too many gauge
symmetries. But after explicit calculation, one finds that the Higgs/sneutrino
potential is minimized at the origin with $V=0$, so
no SUSY-breaking occurs. It thus becomes quickly apparent that some new
fields must be added to do the job of SUSY-breaking.

By putting in a set of new fields, it is easy enough to break SUSY
(for example, through an O'Raifeartaigh-type superpotential).
But how, if at all, can these fields couple to the usual MSSM fields?
One of the properties of SUSY which is
preserved after spontaneous breaking~\cite{dg} is the famous
supertrace formula, applied separately to each individual supermultiplet:
\beq
\mbox{STr}\,M^2=0.
\label{eq:str}
\eeq
This formula
is phenomenologically untenable, for it predicts that the masses of scalars
are distributed evenly above and below the fermion masses. For example,
one of the up-type squarks must be no heavier than an up quark!

Eq.~(\ref{eq:str}) is true only at tree-level.
Even so, this constraint requires that a SUSY-breaking sector must not have
renormalizable tree-level couplings to ordinary matter. Because the
SUSY-breaking must be kept at some distance from the SM sector, it has
been called variously a ``hidden'' or ``secluded'' sector, while the SM
is said to live in the ``visible'' sector.

Because the actual breaking of SUSY is far removed from the SM, the question
of how SUSY is broken is also far removed from experimental probes.
Thus, this question becomes subsidiary to the next
issue we will consider, namely that of communicating SUSY-breaking to the SM.
Indeed, of the multitude of models which we now know to break SUSY and which
could live in the hidden sector, it is often the case that the resulting
visible-sector phenomenology is much less sensitive to model-specific details
than to the method itself by which the SUSY-breaking is communicated.
We therefore leave further discussion on the topic of SUSY-breaking
to the contribution of M.~Peskin to this volume.

Note that there is one aspect of SUSY-breaking that
may nevertheless have universal applicability,
even if we do not know the details of the
SUSY-breaking sector itself. Within a field theory, SUSY-breaking generically
occurs because $F$-terms receive VEV's (the case of $D$-term VEV's rarely
drives SUSY-breaking in realistic models). The $F$-VEV's are controlled by
the superpotential as in the original O'Raifeartaigh model, which always
requires
that a mass scale be present to set the scale of the VEV's. In order to
obtain weak-scale SUSY, we expect that this
mass scale, regardless of the method of SUSY-breaking, will be far below the
Planck scale. Is this natural {\it \`a la}\/ Dirac?

Witten realized that in fact this {\it is}\/ natural if the scale in the
superpotential comes from strong-coupling dynamics in some asymptotically
free gauge theory~\cite{w}. Indeed, in strongly coupled gauge theories,
potentials of the form
\beq
V=\frac{\Lambda^n}{\varphi^{n-4}}+\lambda\varphi^4
\eeq
are typical:  the first term might arise from instantons, gaugino condensation,
or
other strong dynamics, while the second term occurs at tree-level. Such
a potential breaks SUSY, with $F_\varphi\sim\Lambda^2$. Furthermore, since
$\Lambda$ is the strong-coupling scale of the gauge group, it can be
expressed via dimensional transmutation as
\beq
\Lambda=\mpl\, e^{8\pi^2/bg^2(\mpl)}
\eeq
where the one-loop $\beta$-function coefficient $b$ is negative.
Thus $\Lambda$ is a new scale in the theory, exponentially far from the
Planck scale. In this way, SUSY-breaking may hold the key to understanding
the fundamental hierarchy problem of particle physics, explaining why the
Planck
scale is so far from the weak scale.

\section*{\fbox{Question \#7}~~  Once SUSY is broken, how do we find out?}

We have said that the dynamics of SUSY-breaking must occur far from the sector
of the SM. But this begs the question: how does the SM ``learn''
that SUSY has been broken? Remember that at tree-level,
$\mbox{STr}\,M^2=0$, a constraint that must be violated in the visible
sector. We have already hinted at how this can be done: {\it Eq.~(\ref{eq:str})
is true only at tree-level and for renormalizable interactions}.
Two routes therefore seem open to
us for communicating SUSY-breaking to the visible sector: loops and
non-renormalizable interactions. Both have found application in realistic
models.

\subsection*{7.1~~Supergravity mediation}

The ``default'' mechanism for communicating SUSY-breaking is through the
non-renormalizable interactions found in supergravity theories.
(A full introduction to supergravity as a mediator of SUSY-breaking can be
found in
the contribution of R.~Arnowitt and P.~Nath to this volume, or in the standard
references~\cite{mssm}.) Local
SUSY, {\it i.e.}\/, supergravity, is automatically a non-renormalizable field
theory for gravity, containing the spin-two graviton and its spin-3/2 partner,
the gravitino. The Lagrangian for a supergravity model is determined in
terms of three arbitrary functions of the superfields: the superpotential
$W(\varphi)$, the K\"ahler potential $K(\varphi,\varphi^\dagger)$,
and the gauge kinetic function $f(\varphi)$. Note that
$W$ and $f$ are holomorphic
in $\varphi$, while $K$ is real.

The minimal supergravity-mediation model relies on the following assumptions:
\begin{itemize}

\item The superpotential can be written in the form $W=W_H(X)+W_V(\varphi)$
where $W_H(X)$ is the superpotential for the hidden sector fields
$X$,
and $W_V(\varphi)$ is the superpotential for the visible sector fields
$\varphi$.

\item The K\"ahler potential is the minimal one: $K=\sum_i X_i^\dagger
X_i + \sum_i \varphi_i^\dagger \varphi_i$.

\item The gauge kinetic function is given as $f=cX/\mpl$ for some
constant $c\sim\order(1)$.

\item In the hidden sector, SUSY breaks such that $\vev{F_X}\neq0$,\break
$\vev{W_H}\sim\vev{F_X}\mpl$, and $\vev{V}=0$.
\end{itemize}
After SUSY-breaking, the scalar potential of supergravity
\beq
V(\chi)=
e^{K/\mpl^2}\left(
\left|\frac{\partial W}{\partial\chi_i}+\frac{\partial K}{\partial
\chi_i}\,\frac{W}{\mpl^2}\right|^2-3\frac{|W|^2}{\mpl^2}\right)
   ~~~{\rm for}~\chi=X,\varphi
\eeq
reduces in the visible sector to
\beq
V(\varphi)=e^{\vev{K}/\mpl^2}\,\frac{|\vev{W_H}|^2}{\mpl^4}\sum_i|\varphi_i|^2~,
\eeq
giving masses to all visible sector fields $\sim F_X/\mpl$. One of the
more well-known features of this mechanism is that all the visible-sector
scalars receive exactly the same mass, leading to the well-advertised mass
universalities
of supergravity. Similar analyses give universal trilinear and bilinear
terms, also $\sim F_X/\mpl$. The gaugino masses arise from
\beq
{\cal L}=\frac{1}{4}  \,e^{K/2\mpl^2} \, \sum_i\, \frac{\partial f_a}
{\partial\chi_i}\mpl\left(\chi_i
+\frac{\mpl^2}{W}\frac{\partial W}{\partial\chi_i}\right)\lambda^a\lambda^a~,
\eeq
again yielding (universal) masses $\sim F_X/\mpl$.

If supergravity is the {\it dominant}\/ mediator of
SUSY-breaking, then the weak scale can be {\it defined}\/ to be $F_X/\mpl$,
\ie,
$F_X\sim\mz\mpl$. In fact, even if supergravity is {\it not}\/ the dominant
mediator, it will still communicate SUSY-breaking to the visible sector and
the masses induced will always be $\sim F_X/\mpl$. Many of these results can
be summarized neatly by writing the effective operators for the soft masses
in superspace:
\bea
m^2\varphi^*\varphi&\sim&\int d^4\theta\,\frac{X^\dagger X}{M^2}
\varphi^\dagger\varphi=\frac{|F_X|^2}{M^2}\varphi^*\varphi \nonumber \\
M_a\lambda^a\lambda^a&\sim&\int d^2\theta\,\frac{ X}{M}
 W^{\alpha a}
 W^a_\alpha=\frac{F_X}{M}\lambda^a\lambda^a \label{meff} \\
A_{ijk}\varphi_i\varphi_j\varphi_k&\sim&\int d^2\theta\,\frac{ X}{M}
\varphi_i\varphi_j\varphi_k=\frac{F_X}{M}\varphi_i
\varphi_j\varphi_k \nonumber
\eea
where $M$ is the scale of the messenger interactions/particles.
In supergravity, $M=\mpl$. Note that all
masses are of the same order, $m\sim F_X/M$, as we found in supergravity.

The final outputs of minimal supergravity are four mass parameters which
describe all the soft masses of the MSSM: a universal scalar mass $m_0$,
a universal gaugino mass $M_{1/2}$, a universal $A$-term $A_0$, and a
universal $B$-term $B_0$. (In addition, one must also specify the
SUSY-preserving $\mu$-term in order to fully describe the MSSM Lagrangian.)
These four parameters can then be evolved from the
GUT or Planck scale down to the weak scale in order
to form the basis for realistic
SUSY phenomenology studies~\cite{sugramodels}.

It is obvious why such a scenario has been the favored mechanism for
communicating SUSY-breaking since its inception:  it simplifies the
spectrum of the MSSM considerably; it is automatic in the sense that any theory
that connects gravity to a supersymmetric field theory would seemingly have
to include supergravity; and, at lowest order, it produces the kind of
universal masses necessary to solve the FCNC problem described previously.
Why should we even consider anything else?

It is by now well-known that, beyond the lowest order,
there are many effects in supergravity
models that can significantly disrupt the mass
universality at the weak scale. For example, there is no way to forbid
all terms of the form $y_{ij}(X^\dagger,X)
\varphi_i^\dagger\varphi_j$ from appearing in the K\"ahler potential $K$.
Though such terms are suppressed by powers of
$1/\mpl$, the hidden sector fields receive VEV's $\sim\mz\mpl$ so that terms
of this type contribute to the scalar masses with size $\sim\mz$. Furthermore,
because
gravity does not ``know about'' the mass basis choice imposed by the Higgs
Yukawa interactions, there is no reason for $y_{ij}$ to be diagonal
in the same basis as the fermions.
There are also other effects, including RGE running
in the third generation, which can spoil universality and lead to
observable FCNC effects. More generally,
as we have already emphasized,
any source of flavor physics between the Planck and weak
scales will tend to violate the degeneracy.

\subsection*{7.2~~Gauge mediation}

One might hope for some way of communicating SUSY-breaking
that yields mass universality more robustly.
In that vein, there has been much recent interest in so-called
``gauge-mediated'' models~\cite{gmsb}.
The basic principle for these models is rather
simple: If the scalar soft masses are functions only of the gauge charges
of the individual sparticles, universality is automatic. (Remember that
universality in this context only refers to sparticles with identical
quantum numbers, such as
$\widetilde{d}_L,\widetilde{s}_L,\widetilde{b}_L$-squarks.)
Furthermore, if the scale at which
the communication of SUSY-breaking takes place is
well below the Planck scale, then the Planckian ``corrections'' discussed
above cannot disrupt the universality ($F_X/\mpl\ll\mz$).

We will not say much here about the details of gauge-mediation; interested
readers should see the contributions of M.~Peskin and S.~Dimopoulos
to this volume. The effective
mass operators are changed from those in Eq.~(\ref{meff}) in two ways:
first, the messenger scale is now $M\ll\mpl$, and second, the soft masses arise
through loops, so each operator experiences an additional $n$-loop suppression
$\sim (\alpha/\pi)^n$. Specifically, we obtain
\bea
m^2\varphi^*\varphi&\sim&\left(\frac{\alpha}{\pi}\right)^2\frac{|F_X|^2}{M^2}
\varphi^*\varphi \nonumber \\
M_a\lambda^a\lambda^a&\sim&\left(\frac{\alpha}{\pi}\right)\frac{F_X}{M}
\lambda^a\lambda^a \\
A_{ijk}\varphi_i\varphi_j\varphi_k&\sim&\left(\frac{\alpha}{\pi}\right)^2
\frac{F_X}{M} \varphi_i\varphi_j\varphi_k. \nonumber
\eea
If the scalar mass is identified to be $\sim\mz$, then the gaugino mass
will also be $\sim \mz$,
but the $A$-term will be $\sim(\alpha/\pi)\mz\ll\mz$. So
non-zero $A$-terms essentially do not arise in gauge-mediated models, though
they may reappear through renormalization group flow.

Can experiments differentiate this type of mediation from supergravity
mediation?
Perhaps most significantly for phenomenology, these models predict that the
lightest SUSY particle will be the gravitino and that other SUSY particles
can decay into it with observable lifetimes. For an interesting portion of
the parameter space of these models, the missing-energy signal typical
of SUSY models is augmented by two hard photons. However, the rest of the
phenomenology of these models is very similar to that of the supergravity
models.

\subsection*{7.3~~Mediation via pseudo-anomalous $U(1)$}

Finally, there also exists one additional
method for communicating SUSY-breaking that we shall
mention. In string theories, there is often one $U(1)$ gauge group factor
(typically denoted $U(1)_X$) whose
fermionic matter content appears to be anomalous but under which the
string axion field
transforms non-linearly, cancelling the anomaly.  (This will be discussed
in more detail after Question \#18.) The $U(1)_X$ gauge fields
by necessity have
interactions in both the visible {\it and}\/ hidden sectors, interactions which
can communicate SUSY-breaking~\cite{bddp}.
Because of the anomalous matter content, the
$U(1)_X$ gauge superfield acquires a Fayet-Iliopoulos term at one-loop  order
which
breaks the gauge symmetry at a scale one to two orders of magnitude below the
Planck scale: $\epsilon\equiv M/\mpl\simeq 10^{-(1-2)}$.
The effective visible sector mass operators are analogous to those in
Eq.~(\ref{meff}), except that the $X$ fields are
not singlets but are instead charged under the anomalous $U(1)_X$. Thus
the scalar masses can still arise as they do in Eq.~(\ref{meff}), but
the gaugino masses and $A$-terms cannot. For these latter cases, we must
make the replacement
\beq
\int d^2\theta\,\frac{ X}{M} \longrightarrow
\int d^2\theta\,\frac{ X^+ X^-}
{M^2}=\frac{F_{X^+}X^-}{M^2}\sim\epsilon M.
\eeq
Thus the gauginos are generically much lighter than the scalars. In
order to satisfy
experimental bounds on the gauginos, the scalars must then be very heavy
($>1\tev$). This would
reintroduce the naturalness problem of the SM. One solution is that the
third generation scalars have no $U(1)_X$ charge and so they and the gauginos
receive masses $F_X/\mpl\equiv\mz$, while the first and second generation
scalars are
charged under $U(1)_X$, giving them masses $\sim\mz/\epsilon$ where they would
be somewhat immune to the supergravity corrections which could lead to FCNCs.
(There is also the possibility that for $\epsilon$ small enough, such
states could decouple from flavor-changing processes; the problems here would
be the same ones that have been noticed for the decoupling scenarios discussed
after Question \#2.)

The phenomenology of these models has not been explored in any great detail.
Since these models offer the chance to combine the best parts of the
universality and
decoupling solutions to the SUSY flavor problem, they may deserve
more attention.

\vfill
\eject
\separator

\setcounter{footnote}{0}

 ~

\noindent
{\large\bf Section III:~~ Open Questions in Simple Extensions of the
\phantom{Section III:~~}~~ MSSM}

 ~

Having considered the various issues that can arise in the MSSM, we now
turn our attention to two of its simplest and best-motivated extensions.
Probably the simplest extension of any gauge theory is to add to the
spectrum one or more states which are complete gauge singlets. In specific
SUSY models, gauge singlets are often introduced whose VEV's can
provide mass scales without breaking gauge symmetries; the best example is
the case of a singlet solution to the $\mu$-problem already discussed
after Question \#4.  The simplest possible extension of the SM gauge group
is the addition of extra $U(1)$ factors. Such models have been considered
in the past for many reasons:  extra $U(1)$'s arise naturally in
higher-rank GUT groups, they are useful for communicating SUSY-breaking
 as in gauge-mediated models, {\it etc}.  Depending on the model, these
$U(1)$'s can arise in either the hidden or the visible sector, and as we shall
see, each case has its own set of open questions.

\section*{\fbox{Question \#8}~~ Can gauge singlets and SUSY coexist?}

We have already argued that, at least for the $\mu$-problem, it might be
useful to add a gauge singlet to the spectrum of the MSSM.  But there are
two primary barriers to doing so, one cosmological and the other
fundamental. If a singlet $S$ appears in the MSSM coupled to $H_UH_D$ in place
of an explicit $\mu$-term, the action of the MSSM possesses a $\IZ_3$
discrete symmetry under which all superfields are singly charged. When $S$
receives a VEV at the weak scale, it breaks the $\IZ_3$ symmetry and
could, in principle, precipitate the formation of domain walls in the
universe. Such walls would dominate the energy density of the universe,
yielding $\Omega\gg1$. Solutions to this problem usually involve either a
period of late inflation or breaking the $\IZ_3$ symmetry
explicitly~\cite{asw} through non-renormalizable terms in the
superpotential or K\"ahler potential.

However, the more fundamental problem arising for fields which are gauge
and global singlets is that the tadpoles associated with them can
reintroduce quadratic divergences which destabilize the
gauge hierarchy~\cite{singlet}. These tadpoles can arise at one-loop order for
non-minimal K\"ahler potential $K$, or at two-loop order even
if $K$ is minimal. Since tadpole
diagrams arise only for gauge singlets, the loop will be cut off by
the scale of some new physics, usually new physics under which
the singlet accrues gauge charges. Similar arguments can
also be made for global symmetries; in this case, however, we
should demand ``local naturalness'' since gravitational corrections to
the K\"ahler potential
may violate the global symmetry and reintroduce the tadpoles with ${\cal
O}(1)$ coefficients.

As a simple example, let us consider the K\"ahler potential
\beq
   K =  \cdots + (N+N^\dagger)\Phi^\dagger\Phi/\mpl
\eeq
where $N$ is the singlet and $\Phi$ is
any light chiral superfield in the theory. At one-loop order,
the resulting contribution to the
Lagrangian in a theory with supergravity is given by
\bea
\delta{\cal L}&\sim& \frac{1}{16\pi^2}
\frac{\Lambda^2}{\mpl}\int d^4\theta\,e^K(N+N^\dagger) \nonumber\\
&\sim& m_{3/2}^2\mpl N+m_{3/2}\mpl F_N~.
\label{newname}
\eea
Here $\Lambda$ is the
cutoff for the loop integration and can be taken to be the scale at which
the singlet picks up some charge; for true singlets we take
$\Lambda=\mpl$.  The final equality in Eq.~(\ref{newname})
follows from the fact that
the superspace density $e^K$
receives a VEV of the form
$\vev{e^K}\sim1+m_{3/2}\theta^2+m_{3/2}^2\theta^2\overline\theta^2$ in the
process of SUSY-breaking.
Unless the gravitino mass is exceedingly
small (as can happen in some models of low-energy
SUSY-breaking~\cite{np}), this contribution destabilizes the singlet VEV,
pulling it --- and whatever it couples to --- up to large values.
However, it may be possible to build realistic, and very generic,
models of GUT- and intermediate-scale symmetry breaking which are actually
driven
by the tadpole contributions rather than upset by them~\cite{kpp}.
Thus, what may have seemed a problem may indeed become a virtue.

\section*{\fbox{Question \#9}~~ How do extra $U(1)$'s fit into SUSY?}

There are two primary ways of extending the gauge structure of the MSSM:
we can embed the MSSM gauge groups into a large simple group as with GUT models
(see
Question \#13 below), or into a larger direct-product group structure.
In the second case, which we shall discuss here, it is difficult to
build realistic models in which this additional gauge-group structure is
non-abelian,
for such extensions typically require
extending the multiplet structure of the MSSM $SU(3)\times SU(2)$ gauge groups.
On the other hand, additional {\it abelian}\/ gauge groups are relatively
simple to introduce,
and thus they find their way into many possible extensions of the MSSM.
(For a full discussion, see the contribution of M.~Cveti\v c and P.~Langacker
to this volume.)

In non-supersymmetric models, the scale at which the additional gauge
group $U(1)'$ breaks is arbitrary.  This is partially due to the fact that
it is difficult (and perhaps impossible) to stabilize the gauge hierarchy
in such theories. Within the context of supersymmetric theories, however,
the scales of extra gauge interactions are tightly
constrained by the form of the SUSY scalar potential. There are two
primary cases that one can consider.  The first possibility is that the
$U(1)'$ breaks along a direction in the potential
which is $D$- and $F$-flat, so that the scale
of symmetry breaking is set by non-renormalizable operators and/or
radiative corrections to the potential~\cite{cl}.  The second possibility
is that the breaking of the additional $U(1)'$ gauge symmetry
is not along a flat direction.  In this case the symmetry-breaking scale
is constrained by SUSY to be very close to the weak scale.  The first
possibility is
difficult to rule out, and is in fact
highly model-dependent. The second possibility, by contrast, is in many ways
more
natural, but begs the question: if the new interactions should lie near
the weak scale, where are they?

If the extra gauge interactions live in the hidden sector, the argument
against non-abelian groups disappears.  However, the only interesting scale for
symmetry-breaking from the point of view of the visible sector
is the scale at which SUSY is also broken. For
non-abelian groups, any of the previously discussed methods for
communicating SUSY-breaking to the visible sector would now play their
role, and the physics of the hidden sector itself would become difficult to
probe.
But in the case of an extra abelian symmetry, something else can
occur.

If SUSY is broken in the hidden sector at a scale $\Lambda\gg\mz$,
then it is expected that a VEV of size $\sim \Lambda^2$ for the $U(1)$ $D$-term
will be generated.
This in itself is not undesirable. However there is another
generic feature of models with multiple $U(1)$'s which, when combined with
such large $D$-terms, can become dangerous~\cite{dkm}. Since for a $U(1)$
interaction the gauge field strength tensor $F_{\mu\nu}$ is gauge-invariant,
the Lagrangian can contain terms which mix the field strengths of two
different $U(1)$'s.   Specifically, we can have
\beq
         {\cal L}~=~-\frac{1}{4}F^{(a)}_{\mu\nu}F^{(a) \mu\nu}-
	\frac{1}{4}F^{(b)}_{\mu\nu}F^{(b) \mu\nu}-
	\chi \,F^{(a)}_{\mu\nu}F^{(b) \mu\nu}~+~\cdots
\eeq
for a $U(1)_a\times U(1)_b$ theory. Even if $\chi=0$ at tree-level, it can be
generated by loops. And because the mixing operator is dimension-four,
contributions from massive (\eg, stringy) states do not decouple since they are
not suppressed by $\mpl^{-1}$.

When this ``gauge kinetic mixing'' is generalized to the SUSY case, mixing
of the field strengths $F_{\mu\nu}^{(a)}$ implies mixing of the
field strength spinors $W_\alpha$ which in turn implies mixing of their
$D$-components. On integrating
out the auxiliary $D$-fields, the scalar potential of each sector
is sensitive to
the SUSY-breaking $D$-VEV's that are present in the other sector.
Thus the squarks, sleptons, and Higgs bosons of the MSSM, all of which are
charged under
$U(1)_Y$, learn about the SUSY-breaking scale in the hidden sector. Such
contributions, if present, destabilize the gauge hierarchy in the MSSM.

Are there ways out of this disaster? Several options exist~\cite{dkm}.
First, this result is special for extra $U(1)$'s; such gauge kinetic
mixing cannot occur for non-abelian gauge symmetries. Second, there are
discrete symmetries which can forbid such mixing;  these are essentially
charge-conjugation symmetries which act on one $U(1)$ but not the other:
$C(A^{(a)}_\mu)=-A^{(a)}_\mu$ but $C(A^{(b)}_\mu)= +A^{(b)}_\mu$. Such
symmetries can arise naturally if, for example, the two groups are unified
into some non-abelian group $G_N$ whose central $\IZ_N$ is left unbroken
after $G_N\to[U(1)]^2$.

\separator

\setcounter{footnote}{0}

 ~

\noindent
{\large\bf Section IV:~~ Open Questions on SUSY Cosmology}

 ~

The interplay of particle physics and cosmology has never been stronger.
It has always been clear that particle physics provides important inputs
into models of cosmology, but as the field of cosmology has matured, the
opposite has become just as true. SUSY opens exciting new avenues for
cosmology: it can provide the needed dark matter of the universe, it may
provide a natural mechanism for inflation,
it provides several new possibilities for baryogenesis, and
so forth. But the cosmological sword is two-edged, and we find that
cosmology can also serve to constrain SUSY --- for example, a given
particle physics model might overclose the universe, or dissociate light nuclei
after nucleosynthesis, or worse. The next few questions address some of
these important questions concerning the interplay between SUSY and
cosmology.

\section*{\fbox{Question \#10}~~ How does SUSY shed light on dark matter?}

In supersymmetric models with $R$-parity (or matter-parity) conservation,
sparticles can only interact in pairs, thereby guaranteeing that the
lightest SUSY particle (LSP) is absolutely stable. This provides both an
important constraint and the exciting possibility that SUSY may produce
stable, cosmological relics. It is known that non-luminous matter is
needed to explain the rotation curves of galaxies (and galactic clusters)
at large radii where luminous matter densities have fallen near zero. And
even larger densities of ``dark matter'' (\ie, $\rho=\rho_c$ where $\rho_c$ is
the closure density)  are needed in order to place the universe in a stable
evolutionary trajectory such that its current age and density are not
fine-tuned.  Such larger densities are also needed
 in order to reproduce the otherwise successful predictions
of inflation.

However a number of constraints place strong limitations on the form of
the dark matter: nucleosynthesis does not allow very much of the dark
matter to be baryons; heavy isotope searches constrain the ability of
strongly or electromagnetically interacting matter from acting as the
dark matter; and structure formation simulations generally rule out
neutrinos and other particle species which are relativistic when they fall
out of thermal equilibrium. Of all the classes of particles, those which remain
as good candidates for the dark matter are the so-called WIMP's --- Weakly
Interacting Massive Particles.

The MSSM provides two ideal candidates for the dark matter, both of them
WIMP's: the sneutrino and the neutralino. A detailed discussion of how
well each of these particles serves as a dark matter candidate, as well as
a complete list of references, can be found in the contribution of
J.~Wells to this volume; for now let us simply summarize the results.

After detailed calculations~\cite{fos}, one finds that the sneutrino,
though a WIMP, does not provide a good source of dark matter. First, in
most models it is not the LSP. Second, even in those models where it is
the LSP, its relic densities tend to be far from those needed for a dark
matter candidate.  Current experimental bounds from direct detection also
serve to limit the densities of sneutrinos allowed in the solar
neighborhood too severely.

The neutralino can be either a good candidate or a bad one, because the
neutralino is itself an admixture of the bino, wino and the higgsinos,
each with very different properties. Bino neutralinos (\ie, neutralinos
which are mostly composed of binos) are the most common in realistic
models and also provide the best source of dark matter~\cite{dm}. They can
provide reasonable relic densities throughout a broad mass range from tens
to hundreds of GeV (and even thousands of GeV in some parts of parameter
space). Winos, because they interact more strongly, usually provide much
smaller densities for the same range of masses. Higgsinos are in general
poor dark matter candidates.  They tend to interact too strongly and
therefore stay in thermal equilibrium until their densities are
depleted. Even if they could somehow provide the galactic dark matter,
they are very easily detected by a variety of searches. Only if
$\tan\beta$ is very close to one and the Higgsino neutralino decouples
from the $Z$ would the Higgsino neutralino be a good dark matter
candidate~\cite{kw}.

What is most inspiring about the possibility of SUSY dark matter is that
models of particles physics devised solely to satisfy particle physics
constraints and prejudices nevertheless
simultaneously provide a candidate for the
long-sought-after dark matter. (In fact, SUSY had predicted stable relics
even before it was understood that {\it non-baryonic}\/ dark matter was
cosmologically useful.)  Furthermore, it may be possible to study the dark
matter candidate both at accelerators and in dark matter detectors,
hopefully verifying that the properties observed in one match those seen
in the other.  This would close the dark matter question once and for all.

\section*{\fbox{Question \#11}~~ Are gravitinos dangerous to cosmology?}

When global SUSY is broken (at a scale $\sqrt{F}$), there is always a
spin-1/2 goldstino state $\gravitino_\alpha$ in the massless spectrum.
When SUSY is promoted to a local symmetry (supergravity), the goldstino is
eaten by the massless spin-3/2 gravitino.  The resulting fermion has mass
$m_{3/2}\sim F/\mpl$, ``transverse'' components which interact with matter
gravitationally, and ``longitudinal'' components which couple derivatively
to the SUSY current.  As is typical for a Goldstone field~\cite{fayet},
this coupling is suppressed by $1/F$:
\beq
{\cal
L}=\frac{1}{F}\gravitino_\alpha\partial_\mu J_{\alpha\mu}\sim
\bar\lambda^A\gamma^\rho\sigma^{\mu\nu}\partial_\rho\widetilde{G}
F^A_{\mu\nu}
+\bar\psi_L\gamma^\mu\gamma^\nu\partial_\mu\widetilde{G}D_\nu\phi.
\eeq
A lower bound on the gravitino mass is provided by the requirement that
$F\gsim m_Z^2$ so that $m_{3/2}\gsim10^{-5}\ev$; similarly, an upper bound
comes from demanding that $F\lsim\mz\mpl$ so that $m_{3/2}\lsim1\tev$.

In the early universe, gravitinos are believed to have
existed in thermal equilibrium with a
plasma of hidden- and visible-sector fields. As the universe cooled, the
annihilation rate for gravitinos eventually fell below the expansion rate,
and they decoupled, effectively locking in their relic density.

Calculating the relevant cross-sections and solving the Boltzmann equation
allows one to put an upper bound on the mass of a stable gravitino in
order to avoid overclosing the universe. If there exists no mechanism for
diluting the gravitino densities, then one finds~\cite{pp} that
$m_{3/2}\lsim 2h^2\kev$ where $h$ is the Hubble constant in units of
$100\,{\rm km/sec/Mpc}$.  On the other hand, if the gravitino is very,
very light, then its interactions are quite strong and it can stay in
equilibrium below the QCD phase transition.  From standard Big Bang
nucleosynthesis (BBN) results, we know that the number of neutrino species
allowed is $\lsim3$, while a coupled gravitino would behave as an additional
species, violating the bounds. Thus, we find that $M_{3/2}\gsim10^{-6}\ev$ so
that
$\gravitino$ decouples before nucleosynthesis~\cite{lightgrav}. Slightly
stronger bounds (such as $m_{3/2}\gsim10^{-5}\ev$)  can be derived from
limits on the cooling rate for supernova SN1987A via gravitino production
and emission from the core of the star~\cite{sn1987a}.

As with any unwanted relic, the gravitino excess can be diluted away by a
period of inflation. However it is important that the reheating after
inflation not produce a new population of gravitinos. This places upper
bounds on the reheating temperature $T_R$.

Let us consider the case where the gravitino is the LSP (such as in
gauge-mediated models or in no-scale supergravity)~\cite{mmy}. In the mass
range\break $1\kev\lsim m_{3/2} \lsim100\kev$, large densities of
gravitinos will be produced if {\it any}\/ of the MSSM superpartners are
produced, since such superpartners will in time decay to gravitinos. Thus,
assuming that the superpartners are at the weak scale, we have
$T_R\lsim\mz$.  However, if $100\kev\lsim m_{3/2} \lsim(3-300)\gev$ (where
the upper bound depends on the SUSY masses), then the principal source of
gravitinos is not through SUSY decays, but rather through scattering processes
$A+B\to C+\gravitino$; here the reheating temperature must not exceed
$T_R\lsim10^8\,m_{3/2}$. Finally, for $m_{3/2}\gsim (3-300)\gev$, the decay
rate of MSSM superpartners into gravitinos is so small that the decays
take place after nucleosynthesis, disastrously photo-dissociating light
nuclei.

For the first two cases, a period of late inflation (perhaps thermal
inflation --- see Question \#12) can dilute away the gravitino problem.
However, any baryon densities present before this late inflation would
also be diluted away, requiring mechanisms for baryogenesis which operate
at temperatures below $T_R$. This is particularly difficult if
$T_R\lsim\mz$, perhaps requiring electroweak baryogenesis or use of the
Affleck-Dine mechanism.  Note that a period of inflation cannot help for
the last case of $m_{3/2}\gsim(3-300)\gev$.

\section*{\fbox{Question \#12}~~ Are moduli cosmologically dangerous?}

The ``moduli problem'', as we have chosen to call it, is actually a large
class of problems corresponding to the physics of the different moduli
which occur in the MSSM and its extensions.  The essence of the moduli
problem is that fields with extremely flat potentials and weak couplings
to other fields tend to be cosmologically dangerous.

The original example (the ``Polonyi problem'')  is provided by that
hidden-sector field (the ``Polonyi field'') which is a gauge singlet, has a
nearly flat potential and no renormalizable couplings to other matter, and
whose $F$-component is responsible for SUSY-breaking in
supergravity-mediated models.  After receiving a scalar VEV $\sim\mpl$ and
$F$-VEV $\sim\mz\mpl$, the physical Polonyi (scalar) field $\Phi$ emerges
as the scalar partner of the goldstino/gravitino and thus has a mass
$\sim\mz$. The Polonyi VEV, which sets the natural scale of its
oscillations, is much larger than its mass, and so at temperatures far
above the weak scale, $\Phi$ feels only the potential induced by the
(SUSY-breaking) temperature and vacuum energy. In particular, the minimum
in which $\Phi$ finds itself at finite temperature
$T$ and finite Hubble constant $H$
need not correspond to the minimum of the $T=H=0$
potential.

Somewhat more precisely, the problem can be stated like this:  Once $T$
and $H$ fall below $\vev{\Phi}$, $\Phi$ finds itself far from its true
minimum and begins to oscillate with amplitude $\sim\sqrt{\mz\mpl}$.
However, since it is only weakly coupled to matter, there is little
friction to damp the oscillations (\ie, $\Gamma_\Phi$ is small), which
allows the oscillations to continue for times approaching minutes. During
this time, $\Phi$ will dominate the energy density of the universe;  if
$\Phi$ is too long-lived, it will in fact overclose the universe.  But
even if $\Phi$ is not so long-lived, it is the decays of $\Phi$, occurring
long after baryogenesis and perhaps even nucleosynthesis, that are the
principal concern. After several minutes of slow $\Phi$ oscillations, the
temperature of the universe is far below the weak scale, but $\Phi$ decays
are still occurring, dumping entropy into the universe in quantities which
are more than sufficient to overdilute the baryon density (and/or
dissociate light nuclei)  without increasing the temperature enough to
restart the original $B$-violating processes that could replenish it.

The above argument can be generalized and refined by identifying $\Phi$
with other moduli of the theory. These include (but are not limited to):
string moduli whose VEV's $\sim\mpl$ parametrize the size of compact
dimensions;
the string dilaton whose VEV $\sim\mpl$ parametrizes the string coupling
constant; $D$- and $F$-flat directions of the MSSM which have no potential
before SUSY-breaking; and Higgs fields responsible for breaking GUT's to
the MSSM.

A {\it partial}\/ resolution of the moduli problem may rest in the fact
that for many types of moduli, there is no reason for the couplings of the
moduli to other matter to be particularly small.  Thus, once $T,H\lsim
m_\Phi$ and the oscillations begin, $\Gamma_\Phi$ can be sizable, leading
to fast decays which do not appreciably dilute the baryon density. This is
the generalization of one of the early suggestions for solving the Polonyi
problem --- namely, introducing extra fields in the hidden sector to
mediate decays of the Polonyi field into gravitinos~\cite{dfn}. Denoting
$\vev{\Phi}$ at $T=H=0$ as $\vzero$ and at $T,H\neq0$ as $\vth$, we
find that there are then four classes of cosmological histories~\cite{ls}:
\begin{itemize}
\item $\vzero=\vth=0$:  In this case there is no moduli problem and no
	interesting cosmology in the moduli sector.
\item $\vzero=0$ but $\vth\neq0$:  In this case oscillations begin after
	$H\lsim m_\Phi$ but are quickly damped by sizable $\Gamma_\Phi$ so
	there is no moduli problem.  Note that if $\Phi$ carries non-zero
	$B$ or $L$, this case can lead to Affleck-Dine baryogenesis~\cite{ad}.
\item $\vth=0$ but $\vzero\neq0$:  In this case oscillations again begin once
	$H,T\lsim m_\Phi$. However, the moduli cannot have large couplings to
	light particles (otherwise they would not be light), and thus
	$\Gamma_\Phi$ is small and the oscillations last a long time.
        During this time, the universe can
        ``thermally'' inflate~\cite{ls} due to the energy stored in $\Phi$.
	In general such a period of inflation does
	not solve the moduli problem; however, for $\vzero\sim(\mz\mpl)^{1/2}$,
	the problems associated with the inflating moduli do not occur {\it
	and}\/ there is sufficient inflation to dilute any other moduli
	fields.
\item $\vth\neq0$ and $\vzero\neq0$:  In this case the moduli problem
	arises again since the $\Phi$ decays must be suppressed,
	but no period of thermal inflation occurs.
\end{itemize}

For stringy moduli (usually denoted $T$), the conditions necessary to
avoid washing out the baryon density are harder to fulfill. For one thing,
when string moduli begin oscillating, their amplitudes are generally
$\sim\mpl$. Secondly, their couplings to matter are always suppressed by
powers of $m_{3/2}/\mpl$, so that $\Gamma_T$ is very small and the
oscillations last a long time. Thus, more generally than the case discussed
above, one can expect a moduli problem to arise if
$\vev{T}_{T,H}\neq\vev{T}_0$.

If we identify the modulus in question to be the dilaton $S$, many of the
same problems arise, along with a new one~\cite{bs}. Because the dilaton
couples to the vacuum energy density, the non-zero vacuum energies which
are supposed to drive inflation lose much of their energy into driving
oscillations of $S$, thereby slowing the expansion rate. Thus, inflation
must wait until $S$ settles into the minimum of its potential. This itself
is difficult, since the potential for $S$, which arises
non-perturbatively, goes to zero as $S\to\infty$.  Thus, a successful
inflationary model must force $S$ into some local minimum of the full
potential without pushing it over the barrier which separates finite,
time-independent dilaton VEV's from infinite, time-dependent ones. This
appears to be a difficult problem, with no simple solutions currently
available.

There is one other independent mechanism which may help lessen the moduli
problem. If $\Phi$ couples to some other field $\psi$, the equation of
motion for $\psi$ is that of a harmonic oscillator with periodic
driving force, the period being that of the oscillating moduli.  Such an
equation, known as Mathieu's equation, is known to have regions of
instability in which the solutions are exponentially growing. Physically,
these solutions correspond to coherent decays of the moduli at rates far
exceeding those for single particle decays. This phenomenon is known as
parametric resonance~\cite{kls2}.  If parametric resonance occurs, the
moduli will quickly dump most of their energy into particles, instead of
slowing over several minutes.  There are many questions still to be
answered about parametric resonance, the conditions under which it will
occur, and the means by which the decay products thermalize, but this idea
appears to be an exciting advance in our understanding of cosmological
moduli physics.

Finally, it would seem natural to use the moduli themselves as the
inflatons.  We shall not comment on the successes or difficulties in using
any of the moduli, stringy or not, for inflation, but leave that
discussion for the contribution of L.~Randall to this volume.

\separator

\setcounter{footnote}{0}

  ~

\noindent
{\large\bf Section V:~~ Open Questions on SUSY Grand Unification}

  ~

One particularly attractive idea for physics at high energy
scales concerns the possible appearance of a grand unified theory,
or GUT, with a single symmetry group that is large enough to
incorporate the $SU(3)\times SU(2)\times U(1)$ gauge group of the
Standard Model
as a subgroup~\cite{gutidea}.  The idea of grand unification has a long history
independent of SUSY, but once SUSY is included in the picture, a number of
new results and predictions arise. We will consider some of these issues in
this section.

\section*{\fbox{Question \#13}~~ Does the MSSM unify
into a supersymmetric
\phantom{\fbox{Question \#13}~$\,$} GUT?}

There are several profound attractions to
the idea of grand unification.
Perhaps the most obvious is that GUT's have the potential to unify
the diverse set of particle representations and parameters found in
the MSSM into a single, comprehensive, and hopefully predictive framework.
For example, through the GUT symmetry one might hope to explain
the quantum numbers of the fermion spectrum, or even the origins of
fermion mass.
Moreover, by unifying all $U(1)$ generators within a non-abelian theory,
GUT's would also provide an explanation for the quantization of electric
charge;  note that this is a puzzle in the Standard Model due to the
abelian $U(1)_Y$ hypercharge group factor whose allowed eigenvalues are
arbitrary. Furthermore, because they generally lead to baryon-number violation,
GUT's have the potential to explain the cosmological baryon/anti-baryon
asymmetry. By combining GUT's with supersymmetry in the context of
SUSY GUT's~\cite{susyguts}, we then hope to realize
the attractive features of GUT's simultaneously with those of supersymmetry
in a single theory.

\begin{figure}[htb]
\centerline{
   \epsfxsize 3.0 truein \epsfbox {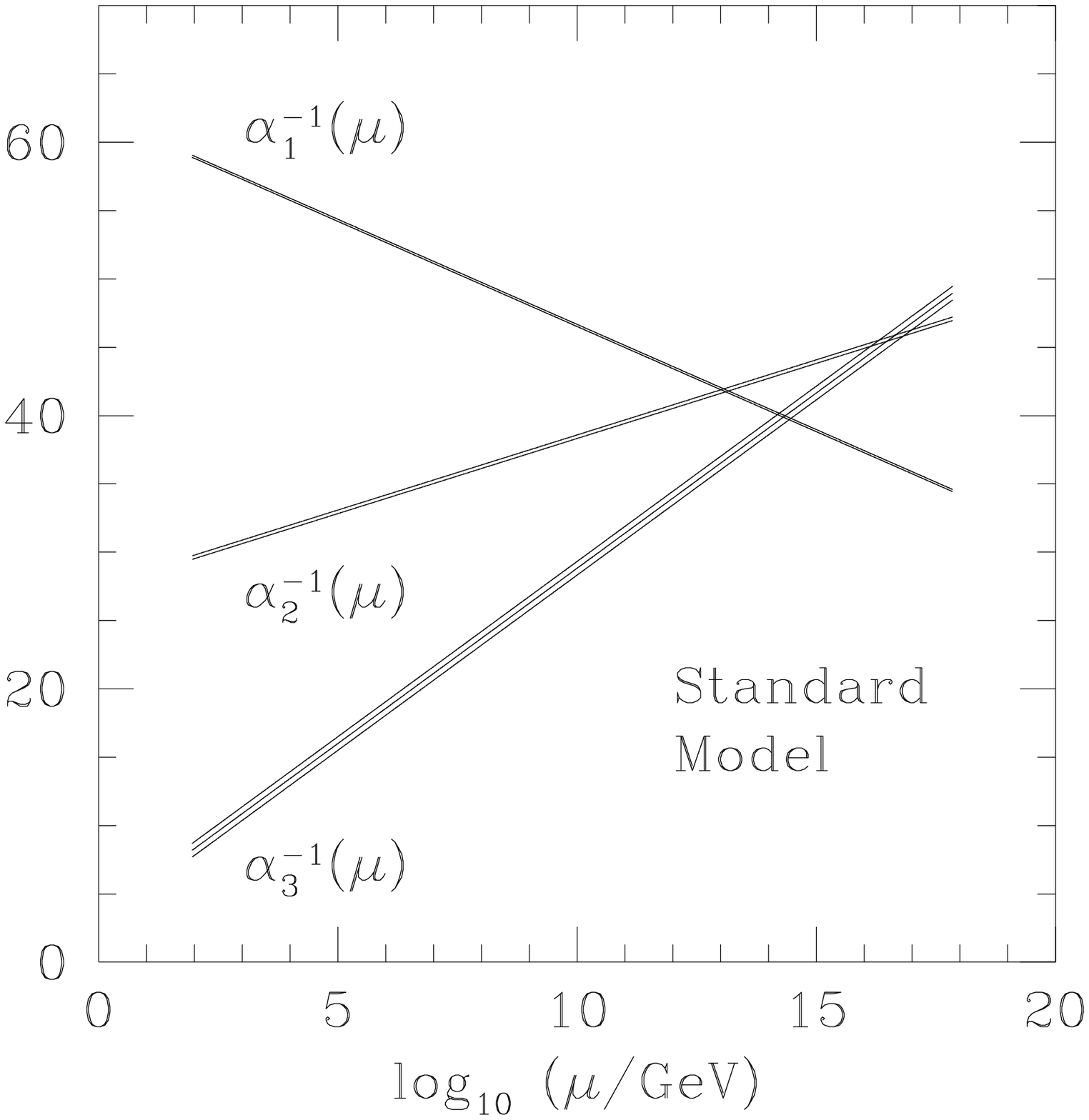}
   \hskip 0.05 truein
   \epsfxsize 3.0 truein \epsfbox {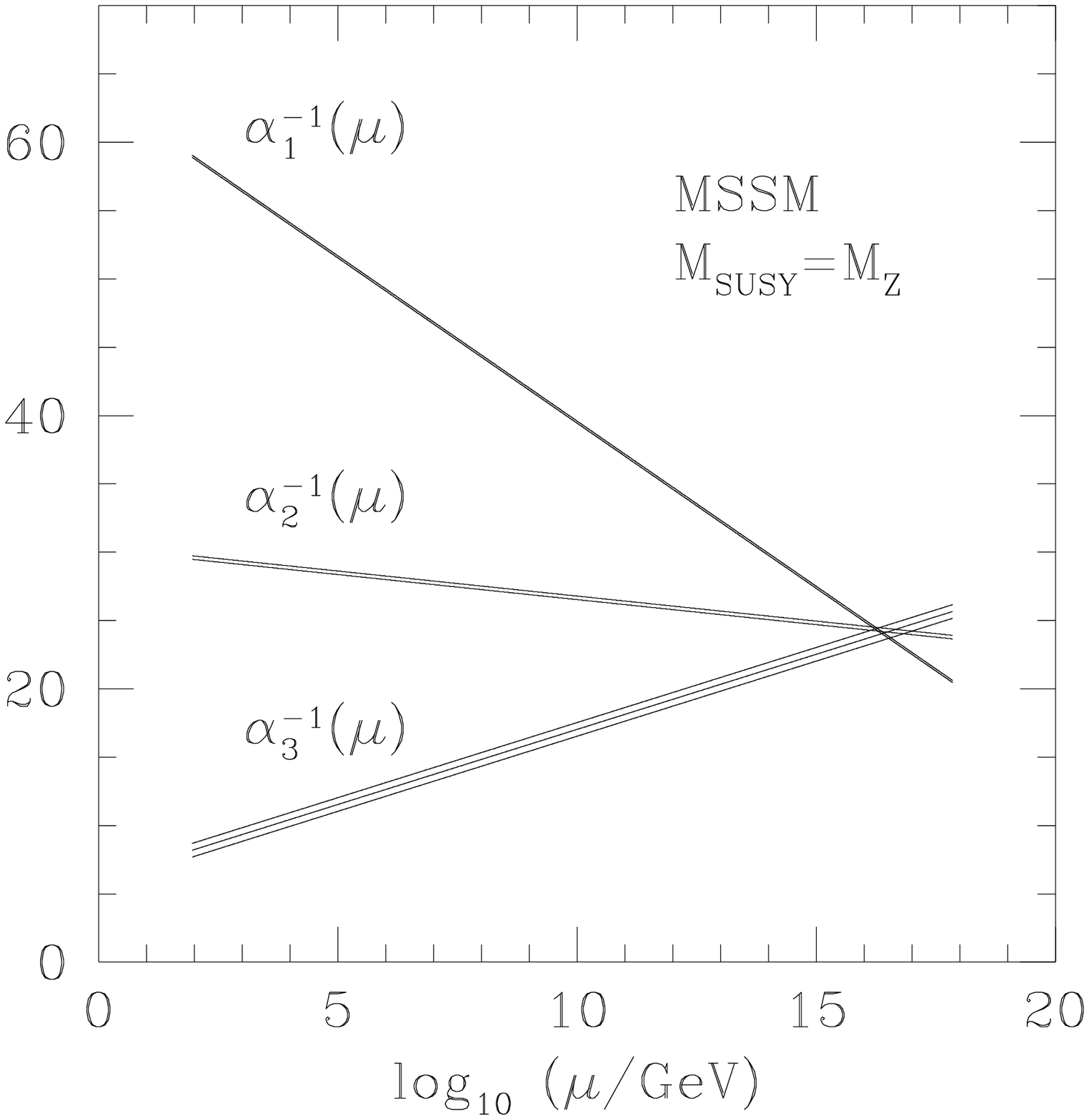}
    }
\caption{One-loop evolution of the gauge couplings within
the non-supersymmetric Standard Model  and within the Minimal Supersymmetric
Standard Model (MSSM).  In both cases $\alpha_1\equiv (5/3)\alpha_Y$
where $\alpha_Y$ is the hypercharge coupling
in the conventional normalization.  The relative width of each line
reflects current experimental uncertainties.}
\label{couplings}
\end{figure}

There is also one compelling piece of {\it experimental}\/ evidence for the
existence of {\it supersymmetric}\/ GUT's.  It is a straightforward matter
to extrapolate the strong, electroweak, and hypercharge gauge couplings
of either the Standard Model or the MSSM to higher energies by using
their one-loop renormalization group equations~\cite{GQW}.
The results are shown in Fig.~\ref{couplings}.
One sees that if this extrapolation is performed within the
non-supersymmetric Standard Model, these couplings fail to unify
at any scale.  However, performing this extrapolation within
the context of the supersymmetric MSSM --- {\it i.e.}\/, assuming
only the minimal MSSM particle content with superpartners near
the $Z$ scale ---
one obtains an apparent
gauge coupling unification~\cite{ejms,modern} of the form
\beq
    {5\over 3}\, \alpha_Y (M_{\rm GUT}) ~=~
    \alpha_2 (M_{\rm GUT}) ~=~
    \alpha_3 (M_{\rm GUT}) ~\approx~ {1\over 25}
\label{gcunification}
\eeq
at the scale $M_{\rm GUT}\approx 2\times 10^{16}$ GeV.
This single unified gauge coupling is then easy to interpret as
that of a single GUT group $G_{\rm GUT}$ which breaks at the scale $M_{\rm
GUT}$
down to $SU(3)\times SU(2)\times U(1)$.
Note that it is the introduction of {\it supersymmetry}\/ which
enables this gauge coupling unification to take place
without any further intermediate-scale structure.

While there are {\it a priori}\/ many choices for such possible groups
$G_{\rm GUT}$, the list can be narrowed down by requiring groups
of rank $\geq 4$ that have complex representations.  The smallest
possibilities are then $SU(5)$, $SU(6)$, $SO(10)$, and $E_6$.
Amongst these choices, $SO(10)$ is particularly attractive because
$SO(10)$ is the smallest simple Lie group for which a single anomaly-free
irreducible representation (namely the spinor ${\bf 16}$ representation)
can accommodate the entire MSSM fermion content of each generation.
Specifically, under the decomposition
$SO(10)\supset SU(5)\times U(1)' \supset SU(3)\times SU(2)\times U(1)_Y\times
U(1)'$,
the \rep{16} representation decomposes as
\beqn
 \rep{16}&\to&
     \rep{10}_{-1} \oplus \overline\rep{5}_{3} \oplus \rep{1}_{-5}\nonumber\\
    &\to&
     \lbrace
       (\rep{3},\rep{2})_{1/6} \oplus (\overline{\rep{3}},1)_{-2/3} \oplus
          (\rep{1},\rep{1})_1\rbrace_{-1} \nonumber\\
     && ~~ \oplus
     \lbrace
       (\overline{\rep{3}},\rep{1})_{1/3} \oplus
       (\rep{1},\rep{2})_{-1/2}
     \rbrace_{3}
     \oplus
       \lbrace (\rep{1},\rep{1})_{0} \rbrace_{-5} ~.
\eeqn
These representations are respectively identified as
the left-handed quark $Q$, the right-handed up quark $u^c_R$, the right-handed
electron $e^c_R$, the right-handed down quark $d^c_R$, the left-handed lepton
$L$,
and the right-handed neutrino $\nu^c_R$.
Note that all Standard Model particles are incorporated, with all of their
correct
quantum numbers, and no extraneous particles are introduced.
Furthermore, such $SU(5)$-based unification scenarios provide a natural
explanation for the normalization factor $5/3$ which appears
in Eq.~(\ref{gcunification}):   this is simply the group-theoretic factor
by which the Standard Model hypercharge generator must be rescaled
in order to join with the non-abelian generators into a
single $SU(5)$ non-abelian multiplet.

The apparent gauge coupling unification of the MSSM is strong
circumstantial evidence in favor of the emergence of a SUSY GUT
near $10^{16}$ GeV.
However, GUT theories
naturally lead to a variety of outstanding questions.
Understanding the answers to these questions therefore provides a window
into high-scale physics.

\section*{\fbox{Question \#14}~~ Proton decay again:  Why doesn't the proton
\phantom{\fbox{Question \#14}~$\,\,$} decay in $10^{32}$ years?}

Perhaps the most important problem that SUSY GUT's must address is
the proton-lifetime problem.  In general, GUT's lead to
a number of processes that can mediate proton decay.
For example, proton decay can be mediated via the off-diagonal $SU(5)$ $X$
gauge bosons that connect quarks to quarks and quarks to leptons.
Such gauge bosons arise, along with the Standard Model gauge bosons,
in the decomposition of the $SU(5)$ adjoint representation;
they transform in the $(\rep{3},\rep{2})_{-5/6}$ and
$(\overline{\rep{3}},\rep{2})_{5/6}$ representations of $SU(3)\times
SU(2)\times U(1)_Y$,
and thus have fractional electric charges $-4/3$ and $-1/3$ respectively.
Because interactions via these $X$ gauge bosons violate baryon-number
($B$) and lepton-number ($L$) symmetries,
such gauge bosons can mediate proton decay via processes such as $p\to \pi^0
e^+$.
However, in supersymmetric GUT's this is not the dominant
source of proton decay because the $X$ gauge boson must have a
mass $M_X\approx M_{\rm GUT}\approx 2\times 10^{16}$ GeV.
This is two orders of magnitude higher than the expected ``unification'' scale
of non-supersymmetric GUT's.
Furthermore, since this decay is gauge-mediated,
the contribution to the branching ratio for proton decay via
this process goes as $\Gamma\sim g^4 m_p^5/M_X^4$ where
$g\approx 0.7$ is the unified gauge coupling and $m_p$ is the proton mass.
The factor $M_X^{-4}$ is a typical suppression for a dimension-six operator,
and results in an expected
lifetime $\tau(p\to \pi^0 e^+) \approx 10^{36}$ years.

A much more problematic dimension-five contribution arises
in supersymmetric GUT's via mediation by colored Higgsino
triplets~\cite{Sakai}.
Because the MSSM requires two electroweak Higgs doublets, and
because a minimal $SU(5)$ GUT gauge structure forces these doublets
to be part of larger ({\it e.g.}\/, five-dimensional)
Higgs representation, the electroweak doublet Higgs will necessarily
have a colored triplet Higgs counterpart which contains a fermionic
colored Higgsino component.
 {\it A priori}\/, a given $SU(5)$-invariant mass term for this
Higgs multiplet will tend to give the same mass to the doublet Higgs
as to the triplet Higgs(ino).
Therefore, since the electroweak doublet Higgs is expected to have
a mass $\sim 100$ GeV, it is generally quite difficult to
give the color triplet Higgs(ino) a large mass.  However, a large mass is
precisely
what we need if we are to avoid rapid proton decay, for this fermionic
Higgsino component of the color Higgs triplet can mediate decay
processes such as $p\to K^+ \overline{\nu}$.  In this case, the
branching ratios go as $\Gamma\sim h^4 m_p^5/M_{\tilde H}^2 M_{\rm SUSY}^2$
where $h\approx 10^{-5}$ is the Higgs(ino) Yukawa coupling to the light
generation
and $M_{\rm SUSY}$ is the scale of SUSY-breaking.
Despite the fact that this branching ratio is Yukawa-suppressed
(by the factor of $h^4$) relative to the dimension-six case,
we have only a factor of $M_{\tilde H}^{-2}$ mass suppression because
the Higgsino mediator is
a fermion. Thus, in order to protect against proton decay
(and also to preserve gauge coupling unification),
the color triplet Higgs must be substantially heavier than
the electroweak doublet Higgs.
Indeed, in order not to violate current experimental bounds,
we must ensure that $\tau(p\to K^+ \overline{\nu})\gsim 10^{32}$ years.
This is the problem of ``doublet-triplet splitting''.
Once the doublets and triplets are somehow split,
supersymmetric non-renormalization theorems should protect
this splitting against radiative corrections.

It is striking that the dominant proton decay mode depends
so crucially on whether or not supersymmetry is present.
Discovery of the  $p\to K^+ \overline{\nu}$
decay mode can thus serve as a clear signal for supersymmetry.

There are a number of potential solutions to the doublet-triplet
splitting problem.  Proposals include the so-called ``sliding
singlet''~\cite{sliding}
and ``missing partner''~\cite{missingpartner}  mechanisms which
apply in the case of $SU(5)$, and also a ``Higgs-as-pseudo-Goldstone''
mechanism~\cite{gift}  which applies in the case of $SU(6)$.
Perhaps the most attractive proposal, however, is the ``missing VEV''
solution for $SO(10)$, originally proposed by Dimopoulos and
Wilczek~\cite{DimWil}.

The basic idea behind this mechanism is as follows.
One way to break the $SO(10)$ GUT gauge symmetry down to that of the MSSM
is to give a vacuum expectation value (VEV) to the adjoint
\rep{45} representation.
However, because the $\rep{45}$ representation contains two Standard-Model
singlets,
there are {\it a priori}\/ many ways in which this can be done
without breaking the Standard-Model gauge group.
The Dimopoulos-Wilczek mechanism entails giving a VEV to only one
of these singlets, and keeping the other VEV fixed at zero.
In order to see this explicitly, it is most useful to consider the
Pati-Salam decomposition under which $SO(10)$ breaks to the Standard Model
gauge group via the pattern
\beq
    SO(10) ~\supset~ SU(4) \times SU(2)_L \times SU(2)_R ~\supset~
                   \lbrace SU(3)\times U(1)_C \rbrace \times
                     SU(2)_L \times U(1)_R~.
\eeq
The hypercharge $U(1)$ is then identified as a linear combination of $U(1)_C$
and $U(1)_R$;
note that $U(1)_C=U(1)_{B-L}$.
Under the first decomposition $SO(10)\supset SU(4)\times SU(2)_L\times
SU(2)_R$,
the $\rep{45}$ representation decomposes as
\beq
    \rep{45} ~\to~
        (\rep{15},\rep{1},\rep{1}) \oplus
        (\rep{6},\rep{2},\rep{2}) \oplus
        (\rep{1},\rep{3},\rep{1}) \oplus
        (\rep{1},\rep{1},\rep{3}) ~.
\label{adjdecomp}
\eeq
However, under $SU(4)\supset SU(3)\times U(1)_C$, we have $\rep{15}\to
   \rep{8}_0 \oplus \rep{3}_{1} \oplus \overline{\rep{3}}_{-1} \oplus
\rep{1}_0$,
while under $SU(2)_R\supset U(1)_R$ we have $\rep{3}\to \lbrace q_C= \pm
1,0\rbrace$.
Thus, only the first and fourth terms in Eq.~(\ref{adjdecomp}) contain
singlets.
The Dimopoulos-Wilczek mechanism consists of giving a VEV to the first singlet
 {\it but not the second}\/.
This works because the $SO(10)$ Higgs decomposes
as $\rep{10}\to (\rep{6},\rep{1},\rep{1}) \oplus (\rep{1},\rep{2},\rep{2})$,
where
the first representation contains the triplet Higgs and the second contains the
doublet Higgs.
By giving a VEV to the $(\rep{15},\rep{1},\rep{1})$ representation within the
$SO(10)$ adjoint but withholding it from the $(\rep{1},\rep{1},\rep{3})$
representation, we see that the effective superpotential term
$\Phi_{\rep{10}} \Phi_{\rep{45}} \Phi'_{\rep{10}}$
gives a mass to the triplet Higgs but not the doublet Higgs.
Constructing a fully consistent $SO(10)$ model in which this mechanism
is implemented in a natural way remains an active area of research,
and many proposals exist~\cite{DimWil,so10modelstwo,so10modelsthree}.

\section*{\fbox{Question \#15}~~ Can SUSY GUT's explain the masses of
\phantom{\fbox{Question \#15}~~~~} fermions?}

In general, the GUT structure imposes not only a unification
of gauge couplings, but also a unification of Yukawa couplings.
Thus, fermion masses are another generic issue that SUSY GUT's must address.
How, through a SUSY GUT, can we explain in a simple way the
many free MSSM parameters that describe the fermion masses?

Just as we did for the gauge couplings, it is straightforward
to use one-loop renormalization group equations (RGE's)
along with the Yukawa couplings
in order to extrapolate the observed fermion masses up to the GUT scale.
In terms of the generic fermion Yukawa couplings $\lambda_i$,
we then find the approximate relations at the GUT scale
\beqn
    &&  \lambda_d(M_{\rm GUT})\approx
      {3}  \lambda_e(M_{\rm GUT})~,
     ~~~
      \lambda_s(M_{\rm GUT})\approx
      {1\over 3}  \lambda_\mu(M_{\rm GUT})~,\nonumber\\
    && \lambda_b(M_{\rm GUT})\approx
         \lambda_\tau(M_{\rm GUT})~.
\label{GUTmassrelations}
\eeqn
Note that because the fundamental GUT idea relates quarks and leptons
within a single multiplet, we are particularly interested in such
mass relations between quarks and leptons~\cite{BEGN}.
The issue, then, is to ``explain'' these relations within the context
of a consistent GUT model.
Ideally, we would also like to explain additional features of the fermion
mass spectrum, such as the inter-generation mass hierarchy,
the masses of the {\it up}\/-type quarks, and
the (near?)-masslessness of the neutrinos.
Reviews of the fermion mass problem within GUT scenarios
can be found in Ref.~\cite{rabyreview}.

Certain features are easy to explain.  For example, the factors
of three that appear in Eq.~(\ref{GUTmassrelations}) can be understood,
as first suggested by Georgi and Jarlskog~\cite{GJ}, as
Clebsch-Gordon coefficients
of the GUT gauge group
(which in turn ultimately stem from the
fact that there are three quarks for every lepton).
This requires the appearance of certain {\it textures}\/
({\it i.e.}\/, patterns of zero and non-zero entries) in the fermion mass
matrices.
Likewise, the inter-generation mass hierarchy might be explained
if the first-generation mass terms are of higher dimension
in the effective superpotential
than those of the second and third generations.
Indeed, within $SO(10)$, even small neutrino masses
can be accommodated via the see-saw mechanism~\cite{seesaw}.

The goal, however, is to realize all of these mechanisms simultaneously
within the context of a self-consistent supersymmetric GUT model.
There are many ways in which such mechanisms can be implemented.
For example, judicious use of a $\rep{126}$ representation of $SO(10)$
can give rise to a heavy Majorana right-handed neutrino
mass, the proper Georgi-Jarlskog factors of three in the light
quark/lepton mass ratios, and GUT symmetry-breaking with
automatic $R$-parity conservation.
Use of the $\rep{120}$ and $\rep{144}$ representations can
also accomplish some (but not all) of the same goals.
General studies of the classes of operators that can explain
the fermion masses can be found in Ref.~\cite{operatoranalysis},
and recent GUT models in which such mechanisms are employed
can be found in Refs.~\cite{so10modelstwo,so10modelsthree,so10modelsone,babu}.

Recently, much attention has focused on deriving GUT models that
are consistent with the additional constraints that come from string theory.
String theory, in particular, tends to severely restrict not only the GUT
representations that might be available for model-building, but also
their couplings (see, {\it e.g.}\/, Refs.~\cite{eln,unif4}).
It turns out that large representations are often entirely excluded,
and only very minimal sets of representations and couplings are allowed.
Recent attempts to build field-theoretic models that are consistent with
these sorts of constraints can be found in Ref.~\cite{so10modelsthree}.
We shall discuss the recent progress in string GUT model-building
in Question \#20.

\separator

\setcounter{footnote}{0}

 ~

\noindent
 {\large\bf Section VI:~~ SUSY Duality}

 ~

Another set of exciting recent developments, made possible
in large part due to supersymmetry, concerns the notion of
SUSY gauge theory {\it duality}\/.
Duality in this context refers to the fact that two
seemingly dissimilar theories can actually describe the
same physics.
A wide variety of exact dualities are known to occur for
$N\geq2$ extended supersymmetries,
but these are somewhat removed from our immediate world
which seems to have at most $N=1$ SUSY. Therefore we will confine ourselves
in this section
to a short discussion of $N=1$ duality.

\section*{\fbox{Question \#16}~~ N=1 SUSY duality:  How has SUSY changed
\phantom{\fbox{Question \#16}~$\,\,$} our view of gauge theory?}

$N=1$ dualities relate two seemingly different theories
in the sense that both flow to the same fixed point in the infrared.
Such theories may be either both asymptotically free, or
one asymptotically free and the other infrared-free.
Supersymmetry plays a fundamental role in uncovering
these duality relations.
By gathering all possible interaction terms into a superpotential
that must be holomorphic in the chiral superfields as well as in their
couplings, supersymmetry imposes extraordinarily tight constraints
on the possible forms of the effective superpotentials that are generated
both perturbatively and {\it non-perturbatively}\/ as one flows
from higher to lower energy scales.  Indeed, in many cases one is
able to determine the effective superpotential exactly.
These exact expressions for the effective superpotentials have been
used for many purposes:  to give new,  simpler proofs of some standard
supersymmetric non-renormalization theorems that hold
beyond perturbation theory;  to determine the situations under which
strong-coupling dynamics can break supersymmetry;  and also
to uncover the phase structure of supersymmetric gauge theories.

$N=1$ dualities come into play when describing the results of this phase
structure analysis.
As a simple example, let us consider $N=1$ supersymmetric $SU(N_c)$
gauge theory with $N_f$ flavors transforming in the fundamental representation.
Such a theory is asymptotically free if $N_f < 3 N_c$;  note that this
is the supersymmetric generalization of the famous constraint $ N_f < (11/2)
N_c$
which holds for non-supersymmetric $SU(N_c)$ theories.
Using the powerful constraints imposed
by the $N=1$ supersymmetry, the infrared limit of this theory has been
determined as a function of the parameters $N_c$ and $N_f$.  One finds
that if $N_f\geq 3 N_c$, the theory flows in the infrared to a (free) theory
of non-interacting quarks and gluons, while if $N_f$ is in the range
$(3/2)N_c \leq N_f\leq 3 N_c$, then the theory flows to a non-trivial
interacting
fixed point.
But what is the infrared limit of the theory if $N_f<(3/2)N_c$?
Evidence suggests that if $N_f> N_c+2$, the infrared limit in this case
is the {\it same}\/ as that for the $N_f \geq (3/2)N_c$ case:  one again has a
free theory of non-interacting elementary constituents.  Indeed, the entire
phase diagram seems to have a symmetry under $N_c\to N_c' \equiv N_f - N_c$,
so that we may identify
\beq
          SU(N_c)~~~{\rm with}~ N_f ~{\rm flavors}
        ~~~\Longleftrightarrow~~~
          SU(N_f-N_c)~~~{\rm with}~ N_f ~{\rm flavors}~
\label{duality}
\eeq
as ``dual'' theories.  The elementary constituents of the infrared limit
of one theory are then identified as the ``dual quarks'' of its dual
theory, and so forth.
It is remarkable that two very different theories can be related in this way.
In fact, this is only the first in a long list of such duality relations, and
examples exist for many other gauge groups and matter representations
(including,
most interestingly, duals between chiral and non-chiral theories).
Anomaly matching conditions provide highly non-trivial checks of these
duality conjectures.
Recent reviews of this subject can be found in Ref.~\cite{IS}.

The existence of such duality conjectures immediately prompts a number of
outstanding
questions.  First, can one {\it prove}\/ these conjectures?
Such a proof would seem to require the construction of a procedure for passing
from
a given gauge theory to its dual, and perhaps also include an explicit mapping
from the degrees of freedom of one theory to the degrees of freedom of the
other.
Second, what does the existence of such dualities tell us about the fundamental
nature of supersymmetric gauge theories?  These dualities suggest, for example,
that the particular gauge symmetry
itself may not be the crucial defining characteristic of such theories.
Finally, of more practical relevance, however, is a third question:  to what
extent do
these duality relations survive supersymmetry breaking?
This will ultimately determine the extent to which such duality conjectures
may be useful for low-energy phenomenology.

Finally, we mention that there exist other sorts of dualities
which also rely heavily on the presence of supersymmetry.  Perhaps the
best-known
among these is Montonen-Olive duality~\cite{OM}, which is an exact strong/weak
coupling duality of finite, interacting, non-abelian gauge theories.
At the present time, this proposed duality can be understood only in the
context
of $N=4$ supersymmetric field theories and finite $N=2$ supersymmetric
theories,
although there do exist extensions to asymptotically free $N=1$ theories.
Once again, a crucial question is whether this duality has an analogue
in (or implication for) non-supersymmetric theories.
It is fair to say that we are only at the beginning stages of understanding for
all of these dualities, and their precise connections and interpretations await
further developments.

\separator

\setcounter{footnote}{0}

 ~

\noindent
{\large\bf Section VII:~~ Open Questions on SUSY and String Theory}

 ~

In this final section, we will discuss some of the phenomenological
connections between supersymmetry and string theory.
We will only focus on general themes and basic
introductory ideas, since many details will be provided in the
subsequent chapters.  For an overall introduction to string theory,
we recommend Ref.~\cite{stringreview}.
There are also a number of review articles that deal with
more specific aspects of string theory.
For example, recent discussions concerning
string phenomenology and string model-building can be found
in Ref.~\cite{phenreview}.
Likewise, reviews of methods of supersymmetry-breaking in
string theory can be found in Ref.~\cite{susybreakreview},
and a review of gauge coupling unification in string
theory can be found in Ref.~\cite{review}.
Note that in these sections we will {\it not}\/ be discussing
some of the more formal aspects of string theory such as string duality;
reviews of this topic can be found in Ref.~\cite{duality}.

\section*{\fbox{Question \#17:}~~ Why strings?}

One of the primary goals of high-energy physics in this century
has been to unify the different observed forces and particles
within the framework of a single, comprehensive theory.
In recent years,
this goal has given rise to tremendous interest in
{\it string theory}\/.  The fundamental tenet
underlying string theory is that all of the known elementary
particles and gauge bosons can be realized as the different excitation
modes of a single fundamental closed string of size $\sim 10^{-33}$ cm.
Thus, within string theory, the physics of zero-dimensional points is replaced
by
the physics of one-dimensional strings, and likewise the spacetime physics of
one-dimensional worldlines is replaced by the physics of two-dimensional
 worldsheets.

There are several profound attractions to this idea.
First, one finds that among the excitations of such a string
there exists a spin-two massless excitation that is naturally
identified as the graviton.  Thus string theory is a theory
of quantized gravity.  Indeed, it is this identification which sets
the fundamental scale for string theory to be the Planck scale.
Second, it turns out that string theory enjoys a measure of finiteness
that is not found in ordinary point-particle field theories,
and therefore many of the divergences associated with field theory
are absent in string theory.
Third, it is found that string theory in some sense {\it requires}\/
gauge symmetry for its internal consistency, and moreover predicts
gauge coupling unification.
But for the purposes of this review, it turns out that the most
intriguing aspect of string theory may be that it seems to predict
supersymmetry.

\section*{\fbox{Question \#18:}~~ What roles does SUSY play in string theory?}

The histories of string theory and supersymmetry are closely intertwined.
Indeed, a form of supersymmetry itself was originally discovered~\cite{Ramond}
in the context of explaining how strings can have fermionic excitations.
We shall now briefly sketch several remarkable inter-relations between
supersymmetry and string theory, focusing on those special roles that
supersymmetry plays in string theory.
We shall mostly restrict our attention to {\it perturbative}\/ string theory,
as this is far better understood than recent developments in possible
non-perturbative formulations of string theory.

\subsection*{18.1~~Worldsheet SUSY, spacetime SUSY, and the dimension of
spacetime}

Perhaps the most intriguing aspect of supersymmetry in string theory
concerns the connections between {\it worldsheet}\/  supersymmetry,
 {\it spacetime}\/ supersymmetry, and the dimension of spacetime.
In general, a closed one-dimensional string sweeps out a
two-dimensional worldsheet with coordinates $(\sigma_1,\sigma_2)$,
and the simplest action for such a string is given by
\beq    S ~=~ \int \,d^2\sigma \, g_{\mu\nu} \,\partial_\alpha X^\mu(\sigma) \,
                 \partial^\alpha X^\nu(\sigma)~.
\label{action}
\eeq
Here $X^\mu(\sigma)$ indicate the spacetime coordinates of the
string as a function of its worldsheet coordinates,
the derivatives are with respect to the worldsheet coordinates,
and $g_{\mu\nu}$ is the spacetime metric.
The spacetime indices run over the range $\mu,\nu=1,...,D_c$.
 From a spacetime perspective, this action is equivalent to the area
of the worldsheet embedded in a $D_c$-dimensional spacetime.
 From the worldsheet perspective, by contrast, this is the action
of a two-dimensional field theory in which the coordinates $X^\mu$ appear
as a collection of $D$ bosonic worldsheet fields with
couplings $g_{\mu\nu}$.
Each different excitation state of the string is then interpreted
as a different particle in spacetime.
Since the fundamental string energy scale is the Planck scale,
only the lowest-lying (massless) excitations are observable,
and the remaining states are all at the Planck scale.

Eq.~(\ref{action}) is the action of the bosonic string,
and it turns out that the quantum consistency of this two-dimensional
action requires that $D_c=26$.
All of the states of this string have integer
spin in spacetime, and are therefore bosons.
However, in order to introduce spacetime fermions,
a natural idea is to supersymmetrize this worldsheet action,
introducing superpartner fermionic fields $\psi^\mu(\sigma)$
on the worldsheet,
\beq    S ~=~ \int \,d^2\sigma \, g_{\mu\nu} \left\lbrack
      \,\partial_\alpha X^\mu(\sigma) \, \partial^\alpha X^\nu(\sigma)
     ~-~i\,\overline{\psi}^\mu \,\rho_\alpha \partial^\alpha\, \psi^\nu
\right\rbrack~,
\label{newaction}
\eeq
where $\rho_\alpha$ are the corresponding {\it two-dimensional}\/ Dirac
matrices.
We then gauge this worldsheet supersymmetry.
This procedure yields the action of the superstring, and
a slight variation on this idea
(one involving a mixture of both supersymmetrized
and non-supersymmetrized actions) yields the action of the heterotic string.
However, in either case, one finds that the spectra of these theories contain
spacetime fermions as well as bosons.
Moreover, the spacetime dimension required for the quantum consistency
of this theory falls to $D_c=10$.

In fact, it turns out that there is an additional remarkable result.
If, in addition to the above gauged worldsheet supersymmetry,
we introduce an additional {\it global}\/ worldsheet supersymmetry
subject to certain constraints~\cite{DixHar},
then the spacetime spectrum of the string not only consists of
bosons and fermions, but actually is itself $N=1$ supersymmetric!
Thus in string theory, $N=1$ supersymmetry in spacetime is realized
as the {\it consequence of two supersymmetries on the worldsheet,
one local and one global}\/!
This is a profound observation, implying that $N=1$ supersymmetry
in spacetime can emerge as (and thereby be explained as) the result of
a more fundamental {\it worldsheet}\/ symmetry (in this case,
$N=2$ worldsheet supersymmetry).
This is only the first of a number of such profound connections
between worldsheet and spacetime supersymmetries, as indicated
in Table~\ref{susytable}.

\begin{table}[htb]
\centerline{
\begin{tabular}{|c|c||c|c|}
\hline
     $N_g$ & $N_t$ & $D_c$ & spectrum \\
\hline
\hline
   0 & 0 & 26 & bosons only \\
\hline
   1 & 1 & 10 & bosons and fermions \\
   1 & 2 & 10 & N=1 SUSY \\
   1 & 4 & 10 & N=2 SUSY \\
\hline
   2 & 2 & 2$^\ast$ & \\
   4 & 4 & $-2$ & \\
\hline
\end{tabular}
  }
\caption{Relations between the total number
  of worldsheet supersymmetries ($N_t$), the number
  of gauged worldsheet supersymmetries ($N_g$),
  the resulting critical spacetime dimension $D_c$ before compactification,
  and the properties of the resulting spacetime spectrum.
  The asterisk indicates {\it complex}\/ dimensions.}
\label{susytable}
\end{table}


\subsection*{18.2~~Supersymmetry, strings, and vacuum stability}

Another intriguing connection between supersymmetry and string theory
concerns vacuum stability.  In field theory, supersymmetry is a very attractive
feature, but it is certainly not {\it required}\/ for consistency.
For example, while the non-supersymmetric Standard Model may
suffer from a variety of unappealing technical problems (foremost among
them the gauge hierarchy problem), it suffers from no fundamental
inconsistency.
In string theory, however, the situation appears to be entirely different.
In general, string theories with non-supersymmetric spacetime spectra
(henceforth to be referred to as non-supersymmetric strings)
have a non-vanishing one-loop tadpole amplitude for a certain light scalar
state called the {\it dilaton}.
As we discussed in Question \#12, such a light dilaton causes a variety of
phenomenological
and cosmological problems.  However, the existence of such a dilaton tadpole
implies that the dilaton experiences a linear potential --- {\it i.e.}\/, that
the ground state of the string is unstable.  Such
a non-supersymmetric string model is then presumed to flow (in the space of
all possible string models) to another point at which stability is restored
and the one-loop dilaton tadpole is cancelled.
A recent study of this question can be found in Ref.~\cite{blumdienes}.

Spacetime supersymmetry is an elegant way of cancelling this dilaton tadpole.
Although it is not known whether all stable string models must be
supersymmetric,
this fact is commonly assumed.
If this assumption is true, then
supersymmetry plays a more profound role in string theory than it does even
in field theory, for the fundamental consistency of the string theory would
seem to require it.
This might then be the best explanation for ``why'' the world
should be supersymmetric, at least at sufficiently high energies.
However, as we stated, it is not known whether this assumption
is true, and we shall see below that certain non-supersymmetric string models
also manage to have a remarkable degree of finiteness and stability.

\subsection*{18.3~~SUSY and pseudo-anomalous $U(1)$'s}

A closely related issue, one with deep ramifications for string phenomenology,
concerns the connection between spacetime supersymmetry and the extra
``pseudo-anomalous'' gauge symmetries that often appear in realistic string
models.

In field theory, consistency requires that there
be no anomalies, and indeed all triangle anomalies are cancelled
in the Standard Model and its supersymmetric extensions.
In string theory, by contrast, there can be $U(1)$ gauge symmetries
(typically denoted $U(1)_X$)
which are ``pseudo-anomalous''.  This means that
${\rm Tr}\, Q_X\not=0$ where the trace is evaluated over
the massless (observable) string states.
The reason this is allowed to occur in string theory
is that string theory provides a different mechanism,
the Green-Schwarz mechanism~\cite{GSmech},
which cancels such triangle anomalies even if this trace is non-zero.
The Green-Schwarz mechanism works
by ensuring that any anomalous variation
of the {\it field-theoretic}\/ $U(1)_X$ triangle diagram
is always cancelled by a corresponding
non-trivial $U(1)_X$ transformation of the
string axion field.
This axion field arises generically in string theory
as the pseudo-scalar partner of the dilaton, and couples universally
to all gauge groups.
Thus, the existence of such a mechanism in string theory
implies that
anomaly cancellation in string theory does not require cancellation of
${\rm Tr}\,Q_X$ by itself, and consequently
a given string model can remain non-anomalous even while having
${\rm Tr}\, Q_X\not=0$.
Indeed, this is the generic case for most realistic string models.

What does this have to do with supersymmetry?
It turns out that {\it even though}\/ the anomalies caused
by having ${\rm Tr}\, Q_X\not=0$ are cancelled by the Green-Schwarz
mechanism,
there is still another danger:
such a non-vanishing trace leads to
the breaking of spacetime supersymmetry at one-loop order
through the appearance of a
one-loop Fayet-Iliopoulos $D$-term of the form~\cite{DSWshift}
\beq
         {g_{\rm string}^2 \, {\rm Tr}\, Q_X \over 192 \,\pi^2 }\,
          \mpl^2
\label{Dterm}
\eeq
in the low-energy superpotential.
This in turn destabilizes the string ground state by generating a
dilaton tadpole at the two-loop level,
and signals that our original string theory (or string model) in which
${\rm Tr}\, Q_X\not=0$
cannot be consistent.

The standard solution to this problem is
to give non-vanishing vacuum expectation values (VEV's) to
certain scalar fields $\phi$ in the string
model in such a way that the offending $D$-term in Eq.~(\ref{Dterm}) is
cancelled and spacetime supersymmetry is restored.
In string moduli space,
this procedure is equivalent to moving
to a nearby point at which the string ground
state is stable, and consequently this procedure is
referred to as {\it vacuum shifting}\/.
The specific VEV's that parametrize this vacuum shift
are determined by solving the various $F$- and $D$-term
flatness constraints, and one finds
that they are typically quite small, of the order
$\langle \phi \rangle/M_{\rm string} \sim {\cal O}(1/10)$.

Such vacuum shifting has important consequences
for the phenomenology of the string theory.
For example, vacuum shifting
clearly requires that those scalar fields
receiving VEV's be charged under $U(1)_X$.
Thus, the act of vacuum shifting breaks $U(1)_X$, with the
$U(1)_X$ gauge boson ``eating'' the axion to become massive.
In fact, since the scalars $\phi$ which are charged under $U(1)_X$
are also often charged under other gauge symmetries,
giving VEV's to these scalars typically causes further gauge
symmetry breaking.
Perhaps most importantly, however, vacuum
shifting can generate effective superpotential mass
terms for vector-like states $\Psi$ that would otherwise be massless.
Indeed, upon replacing the scalar fields $\phi$ by their VEV's in
the low-energy superpotential,
one finds that higher-order non-renormalizable couplings
can become lower-order effective mass terms:
\beq
       {1\over M_{\rm string}^{n-1}}\,
        \phi^n \,\overline{\Psi}\Psi ~\to~
       {1\over M_{\rm string}^{n-1}}\,
        \langle\phi\rangle^n \,\overline{\Psi}\Psi ~.
\label{effmassterm}
\eeq
Moreover,
it often turns out that various string selection rules
prohibit these types of effective mass terms from appearing
in the tree-level superpotential
until rather high order.  For example, we often
have $n\gsim 5$ in Eq.~(\ref{effmassterm}).
Since one typically
has $\langle \phi \rangle/M_{\rm string} \sim {\cal O}(1/10)$,
the effective mass terms that are generated
after the vacuum shift are schematically
of the order $\langle \phi\rangle^n / M_{\rm string}^{n-1}\sim
   (1/10)^n M_{\rm string}$.
Thus, we see that vacuum shifting in string theory provides
an economical  mechanism
for generating intermediate mass scales.

It is remarkable that in string theory,
the need to protect supersymmetry
against the effects of pseudo-anomalous $U(1)$'s can have
all of these important effects.
This once again underlines the key idea that supersymmetry
plays a profound role in string theory --- in some ways,
even more profound than the role it plays in field theory.

\section*{\fbox{Question \#19}~~ How is SUSY broken in string theory?}

Given the unique role of supersymmetry in string theory,
and given that our low-energy world is non-supersymmetric,
the next issue that arises is
the means by which supersymmetry can be {\it broken}\/ in string theory.
Although there are many different proposals, these can be grouped
into essentially three methods:
one can break SUSY within perturbative string theory itself
(so that one obtains a non-supersymmetric string);
one can break SUSY within the low-energy effective field theory
derived from a supersymmetric string;
and one can break SUSY via a new scenario
(the Ho\v{r}ava-Witten scenario)
that makes use of certain features of non-perturbative string theory.

\subsection*{19.1~~Within string theory itself}

Perhaps the most direct way of breaking supersymmetry in string theory
is within
the full string theory itself.  Thus, one would obtain
a string that has no spacetime supersymmetry at {\it any}\/ scale,
not even the Planck scale.  As we stated above, such strings
are generally not stable (due to their non-vanishing dilaton tadpoles),
but it is not known whether there might exist a special subset of
non-supersymmetric strings which {\it are}\/ stable.
In many ways, this question is the stringy analogue of the cosmological
constant problem:  how can one find a non-supersymmetric ground state
which preserves a near-exact (if not absolutely exact)
cancellation of the cosmological constant?  Indeed, in string theory
these two questions are actually related in a deep way, and various proposals
exist for solving this problem~\cite{cancellambda}.

Breaking supersymmetry within the string theory itself
can be done in a variety of different ways.  In all cases, however,
the basic idea is to implement a carefully chosen ``twist''
when compactifying the string so that all superpartner
states (including the gravitinos themselves) suffer so-called
``GSO projections'' and are removed from the string spectrum.
In many (but not all) cases, this method is equivalent to the
well-known Scherk-Schwarz mechanism~\cite{scherkschwarz}
in which supersymmetry is broken through the special dependence
that compactified fields have on the coordinates of compactified
dimensions.
This procedure was introduced into string theory in Ref.~\cite{Rohm},
and has since been pursued in a number of
contexts~\cite{missusy,supertraces,antoniadis,others,others2}.

Breaking supersymmetry this way offers a number of distinct advantages.
The most important may be that it {\it preserves the string itself}\/.
Specifically, because this method results in another {\it string}\/ theory,
it preserves the string symmetries (such as modular
invariance) that underlie many of the properties of string theory
(such as finiteness) that we would like to preserve even after SUSY-breaking.
For example,
it has been shown that even though spacetime supersymmetry is broken in
such scenarios,
there is always a hidden ``misaligned supersymmetry''
\cite{missusy} that remains in the string spectrum.
This misaligned supersymmetry tightly constrains the distribution of bosonic
and fermionic states throughout the string spectrum in such a way
that even though SUSY is broken, bosons and fermions nevertheless provide
cancelling contributions to string amplitudes, and certain mass supertraces
continue to vanish~\cite{missusy,supertraces}.
Indeed, the phenomenology of misaligned supersymmetry ensures
that these supertraces cancel not in the usual scale-by-scale manner,
multiplet-by-multiplet, but rather through subtle simultaneous
conspiracies between physics at different energy scales.
This may have important phenomenological applications.

There is also another important phenomenological aspect of such theories.
In some sense, since SUSY is being directly broken at the string scale,
one might suspect that all gravitinos must have Planck-scale masses.
However, this is not the case:  it turns out that one can often ``dial''
the gravitino mass $m_{3/2}$ in such scenarios.
But various string consistency constraints then imply~\cite{antonlewellen}
that such theories will essentially have an extra dimension
whose radius is $R\sim m_{3/2}^{-1}$.
Thus, in such string models,
the existence of a TeV-scale gravitino implies
the existence a TeV-scale extra dimension, which in turn implies
the existence of infinite towers of TeV-scale string states with
TeV-scale mass separations.
The phenomenology of such scenarios is discussed in Ref.~\cite{antoniadis}.

Finally, we remark that even though breaking SUSY through the string itself
does not provide supersymmetry at any scale below the Planck scale,
this need not be in conflict with gauge coupling unification.
A discussion of this point can be found in Ref.~\cite{review}.

\subsection*{19.2~~Within the low-energy effective theory}

The second way of breaking SUSY in string theory is to start with
a supersymmetric string at the Planck scale, and then break
SUSY within the low-energy effective field theory that is derived
from the massless (observable) modes of the string.
Since this method is essentially field-theoretic in nature,
occurring purely within the language of the effective field theory,
it does not necessarily result in a particle spectrum that can
be interpreted as the low-energy limit of a non-supersymmetric string.
This method therefore presumably breaks some or all of the consistency
constraints that underlie string theory, and destroys the
fundamental finiteness properties of the string.
However, it offers the advantage that a purely field-theoretic
treatment of SUSY-breaking will suffice.

Because this method of SUSY-breaking is field-theoretic, all of the
SUSY-breaking mechanisms we have outlined in Question \#6 apply to this case
as well.  The most commonly assumed scenario
is that the dynamics of extra ``hidden'' string sectors
will break supersymmetry through some mechanism ({\it e.g.}\/,
gaugino condensation~\cite{gauginocondensation}) which is then communicated
to the observable sector through either gravitational or gauge
interactions.  The ensuing phenomenologies are then
analyzed in purely field-theoretic terms, and will be discussed in
upcoming chapters.

\subsection*{19.3~~SUSY-breaking in strongly coupled strings}

Finally, there also exists a third
scenario for SUSY-breaking within string theory.
At strong coupling, it has been proposed~\cite{HW} that the ten-dimensional
$E_8\times E_8$ heterotic string can be described as the compactification
of an {\it eleven}\/-dimensional theory known as `M-theory'
on a line segment of finite length $\rho$.
The two $E_8$ gauge factors
are presumed to exist at opposite endpoints
of this line segment.
In order to incorporate GUT-scale gauge coupling unification within this
scenario, it turns out~\cite{witten}
that the length $\rho$ of the eleventh dimension
must be substantially larger than the eleven-dimensional
Planck length.
Thus, one has a situation in which the two $E_8$ gauge factors
communicate primarily with their own ten-dimensional worlds located at opposite
ends of an eleven-dimensional bulk, and only gravitational interactions
connect these two ``ends of the world'' with each other.

\begin{figure}[htb]
\centerline{
   \epsfxsize 3.5 truein \epsfbox {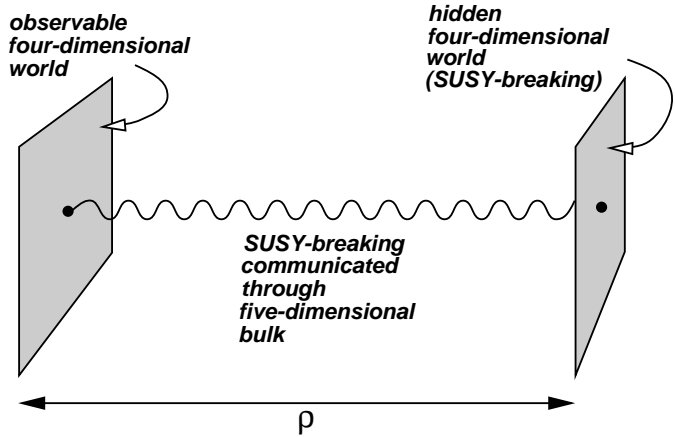}
    }
\caption{The Ho\v{r}ava-Witten scenario for communicating
SUSY-breaking from a hidden world across a five-dimensional
bulk of length $\rho$ to the observable world.}
\label{horwit}
\end{figure}

If we now imagine further compactifying this picture to four dimensions,
we obtain a scenario in which one four-dimensional world is the
``observable'' world that descends from one of the $E_8$ gauge factors,
while the other four-dimensional world
represents the ``hidden'' sector that descends from the other $E_8$ gauge
factor.
Since the radii of the additional six-dimensional compactification
must be considerably smaller than the length of the eleventh dimension,
one obtains an effective situation in which two four-dimensional worlds
are connected through a five-dimensional bulk.
This situation is illustrated in Fig.~\ref{horwit}.
Most interestingly,
if strong-coupling dynamics
in the hidden $E_8$ gauge factor
causes SUSY-breaking to occur in that sector
(such as via gaugino condensation), the effects of such SUSY-breaking
will be communicated gravitationally to the observable world through
the five-dimensional interior bulk.
This scenario thereby places the question of SUSY-breaking
in an entirely new geometric context.  For example, in some circumstances
the resulting gravitino mass can be identified with the radius $\rho$ of the
fifth
dimension.
Various phenomenological consequences of this picture of SUSY-breaking are
currently being explored~\cite{HorWitscenario}, in the context of
both gaugino condensation and Scherk-Schwarz compactification.

\section*{\fbox{Question \#20}~~ Making ends meet: How can we understand\break
\phantom{\fbox{Question \#20}$\,$} gauge coupling unification from string\break
\phantom{\fbox{Question \#20}~~$\,$} theory?}

In this section we will discuss another important issue connected with
strings and supersymmetry, namely gauge coupling unification.
As we have seen, the strong, electroweak, and hypercharge
gauge couplings
appear to unify at approximately $M_{\rm MSSM}\approx 2\times
10^{16}$ GeV when extrapolated within the framework of the MSSM.
Indeed, this observation is often taken
as evidence for supersymmetry, which
also provides elegant solutions for the finiteness and
gauge hierarchy problems.  Thus, the currently accepted
field-theoretic scenario calls for some sort of grand unified
group (GUT) above $M_{\rm MSSM}$;  the MSSM gauge group
and spectrum between $M_{\rm MSSM}$ and
the scale of SUSY-breaking $M_{\rm SUSY}$ at which
the superpartners decouple;
and simply the Standard Model gauge
group and spectrum below $M_{\rm SUSY}$.

This is a compelling picture, except for various problems.
First, $M_{\rm MSSM}$ is close to the Planck scale, but
gravity is not incorporated.  Second, one would in principle
like to explain the spectrum of the MSSM --- \eg, to explain
why there are three generations, or to derive the fermion
mass matrices.  Third, if there is a GUT theory above
$M_{\rm MSSM}$, what about proton-lifetime problems?
One requires some sort of doublet-triplet splitting mechanism,
as was discussed in Question \#14.
Finally, why should we require gauge coupling unification
at all?  This is, after all, only a theoretical prejudice,
and is not required for the consistency of the model.

\subsection*{20.1~~The predictions from string theory}

Of course, string theory can
solve these problems.  First, as we have seen, it naturally
incorporates quantized gravity, in the sense that
a spin-two massless particle (the graviton) always appears
in the string spectrum.  Second, $N=1$ supersymmetric
field theories with non-abelian gauge groups naturally
appear as the limits of a certain class of string models
(the heterotic strings).  Third, string theory
can provide, in principle, a uniform framework for understanding
three generations, fermion matrices, doublet-triplet splitting
mechanism, {\it etc}.\ --- in principle, there are no free
parameters!  Finally,
it also turns out that {\it independently of the existence
of a unified gauge symmetry}, heterotic string theories
always give rise to a natural unification of gauge couplings.
Indeed, in heterotic string theory,
the gauge and gravitational couplings automatically
unify~\cite{Ginsparg} to form a single coupling constant $g_{\rm string}$:
\beq
   8\pi {G_N\over \alpha'} ~=~ g_i^2\,k_i ~=~ g_{\rm string}^2~.
\label{unification}
\eeq
Here $G_N$ is the gravitational (Newton) coupling;
$\alpha'$ is the Regge slope (which sets the mass scale
for string theory); $g_i$ are the gauge couplings;
and the normalization constants $k_i$
are the {\it affine levels}\/  (also sometimes called {\it Ka\v{c}-Moody}\/
levels)
at which the different group factors are realized.
For non-abelian group factors we have $k_i\in\IZ^+$,
while for $U(1)$ gauge factors the $k_i$ are arbitrary.
Thus, string theory appears to give us precisely
the features we want.

There are, however, some crucial differences between
string theory and field theory.
First, string theory is a {\it finite}\/ theory:  the
gauge couplings run only within the framework of the
string low-energy {\it effective}\/ theory.
Second, in string theory all couplings are ultimately
dynamical variables, related to the expectation values
of scalar moduli fields.
The third difference is the dependence on the \KM\
levels $k_i$.  These levels are essentially normalizations,
and are therefore analogous to the hypercharge normalization
$k_Y=5/3$ which appears in $SU(5)$ or $SO(10)$ embeddings,
but in string theory such normalizations also appear
for the {\it non}\/-abelian gauge couplings as well.
It turns out that the most easily constructed string models have $k_i=1$
for non-abelian factors.

The most important difference concerns the scale
of the unification.
The string unification
scale is set by $\alpha'$ (which in turn
is set by Planck scale), and at
the one-loop level one finds~\cite{Kaplunovsky}:
\beq
   M_{\rm string}~\approx ~g_{\rm string}~\times~5\times 10^{17}
       ~{\rm GeV}~.
\label{stringscale}
\eeq
Since extrapolation of low-energy data suggests
that $g_{\rm string}\approx {\cal O}(1)$, we thus find
that $M_{\rm string}\approx  5\times 10^{17}$ GeV ---
a factor of 20 discrepancy relative to the MSSM
prediction!

Is this is a major problem?
A factor of 20 sounds large, but this is only
a 10\% effect in the {\it logarithms}\/ of the mass scales.
Unfortunately, however, this discrepancy
leads to wildly incorrect values for the low-energy observables
$\sin^2\theta_W$ and $\alpha_{\rm strong}$
at the weak scale.  In other words, if we start our MSSM running of
gauge couplings down from
$M_{\rm string}$ rather than from $M_{\rm MSSM}$, we find that
string theory predicts values for these quantities
which differ from their experimentally observed values
by many standard deviations.
This is the problem of gauge coupling unification in string theory.
Essentially, given the high-energy predictions of string theory
and our low-energy experimental couplings, we face the classic question:
how can we make the two ends meet?

\subsection*{20.2~~Overview of possible solutions}

Over the past decade, a number of solutions to this
question have been proposed.
We shall here outline only six possible classes of solutions.
The reader should consult Ref.~\cite{review}  for a more complete discussion of
these
and other solutions.

The first solution reconciles $M_{\rm MSSM}$ and $M_{\rm string}$
by assuming that the three low-energy gauge couplings indeed
unify at $M_{\rm MSSM}$ because of the presence of a
unifying gauge symmetry group $G$ at that scale, whereupon the new unified
gauge
coupling $g_G$ runs upwards to $M_{\rm string}$ where it unifies with the
gravitational coupling.  Thus, at the string scale, we are essentially
realizing the GUT group $G$ as our gauge symmetry:  these are
``string GUT models''.
Note that in this context we therefore consider only
those unification groups $G$ such
as $SU(5)$ or $SO(10)$ which are {\it simple}.
An essential property of such groups
is that they require a Higgs scalar representation in the
adjoint of $G$ in order to
break $G$ down to the MSSM gauge group.

The second possible solution makes use of the \KM\ levels $k_i$
that appear in the string unification relation in Eq.~(\ref{unification}).
Indeed, in string theory, these levels $k_i$ need not take the
values $(k_Y,k_2,k_3)=(5/3,1,1)$ that we na\"\i vely
expect them to have in the MSSM.
It is then possible that
non-standard values for these levels
could alter the runnings in such a way as to reconcile the string
unification scale with the MSSM unification scale.
This would clearly be a stringy effect.

The third solution
supposes that there can be large ``heavy string threshold corrections''
at the string scale.
These corrections represent the contributions from the infinite towers
of massive (Planck-scale) string states that are otherwise neglected
in an analysis of the purely massless string spectrum.  This would
also be an intrinsically stringy effect.

A fourth solution involves ``light SUSY thresholds'' --- the
corrections that arise due to the breaking of supersymmetry --- and
are typically analyzed in field theory.

A fifth solution involves extra matter beyond the MSSM at intermediate
mass scales.
While introducing such matter may seem {\it ad hoc}\/ from the
field-theory perspective, it turns out that certain exotic non-MSSM states
appear in, and are actually {\it required for the self-consistency}\/ of,
many realistic string models.

Finally, a sixth solution~\cite{witten} involves possible
effects due to non-perturbative
string physics.  For example, as we have seen, recent developments
in string duality suggest that at strong coupling, the behavior of
heterotic strings can be modelled by other theories for which the
heterotic string prediction in Eq.~(\ref{unification})
is no longer valid.
This then effectively loosens the tight constraints between the
gauge couplings and the gravitational coupling, which in turn
enables one to separate the gauge-coupling unification scale from
the gravitationally-determined string scale.

Thus, we are faced with one over-riding question:  Which solution(s)
to the problem --- {\it i.e.}\/, which ``path to unification'' ---
does string theory actually take?

It is perhaps worth emphasizing that this a much more difficult question
in string theory than it would be in field theory.  In field theory,
one can imagine rather easily building a model that realizes any one
of the above proposals.  In string theory, however, there are
deeper string consistency constraints which arise
due to the fact
that four-dimensional (spacetime) physics
is ultimately derived from two-dimensional (worldsheet) physics.
Thus four-dimensional
spectra, gauge symmetries, couplings, {\it etc.},
are all ultimately determined or constrained by worldsheet symmetries.
This tends to make it difficult, when string model-building,
to realize one given desirable phenomenological feature
in one sector of a string model
without upsetting some other desired feature in a different sector of
the model.

The question, then, is to determine which of the above
potential solutions to the string unification problem are
self-consistent in string theory, and can be
realized in {\it actual realistic string  models}.

\subsection*{20.3~~Current status}

We shall now give a quick summary of the current status of some of
these proposed solutions.  A more detailed review (along with
appropriate references)  can be found in Ref.~\cite{review}.

 {\it String GUT models}:~~
As mentioned above, the goal in this approach is to construct
realistic string GUT models --- \ie, string models
whose low-energy limits reproduce standard $SU(5)$ or $SO(10)$
unification scenarios.  The major problem that one faces, however,
is that while it is generally easy to obtain the required gauge
group, obtaining the required {\it matter representations}\/
has proven to be very difficult.
The fundamental reason for this difficulty is that:
(i)  the string requirement that the worldsheet conformal field theory
be unitary ends up restricting the allowed massless matter representations
that the string model can produce;  and (ii)  for GUT symmetry breaking,
one requires a Higgs scalar transforming in the adjoint of the
GUT gauge group.  Together, these two requirements imply that
one must realize the GUT gauge symmetry at an \KM\ level
$k_{\rm GUT}\geq 2$, and historically it has proven to be
a highly non-trivial task to construct such a higher-level
string GUT model with three
generations~\cite{stringGUTattempts,aldaz,shynew}.

At present, the three-generation problem has been solved
at level two only in the case of $SU(5)$~\cite{aldaz,shynew,su5models}.
At level three, however, there currently exist three-generation
models for $SU(5)$, $SU(6)$, $SO(10)$, and $E_6$~\cite{KT}.
However, much phenomenological analysis of these models still
remains to be done.
In some cases, these models tend to have extra chiral matter, or unsuitable
couplings.  Doublet-triplet splitting also remains a problem, and
appears to require fine-tuning.
There are also rather tight constraints~\cite{unif4} concerning
the allowed representations and couplings for these models which
restrict their phenomenologies significantly.
However, the important point is that
the issues concerning string GUT model-building now seem to be
more of a technical rather than fundamental nature, and further progress
can be expected.

 {\it Non-Standard Levels and Hypercharge Normalizations:}~~
In this solution to the unification problem,
one attempts to realize the MSSM gauge group and particle content
in a given model, but to
reconcile the discrepancy between $M_{\rm MSSM}$ and $M_{\rm string}$
by having non-standard values for the levels $(k_Y,k_2,k_3)$.
A straightforward analysis~\cite{ibanez,k1paper}
shows that in order to do the job,
the required levels would be:
\beq
      k_2=k_3=1,2~;~~~~~~k_Y/k_2 \approx 1.45 - 1.5~.
\eeq
Thus, restricting our attention the simpler level-one models, the question
arises:
can one even realize realistic string models with $k_Y$ in this range?

This question is motivated by the observation that the standard
$SO(10)$ hypercharge embedding naturally leads to the
MSSM value $k_Y=5/3$,
and most trivial modifications or extensions to this embedding
tend to increase $k_Y$.  Thus, more generally, we ask
whether it is even possible to realize hypercharge embeddings
with $k_Y<5/3$, and whether this would cause undesirable effects
on the rest of such a string model.
Note that one always must have $k_Y\geq 1$ in any string model
containing at least the MSSM spectrum~\cite{Schellekens}.

The current status of this approach is as follows.
In general, it is very difficult to arrange to have $k_Y<5/3$ in string
theory~\cite{k1paper,shynew}.  However, some self-consistent string models with
$k_Y<5/3$ have been constructed~\cite{shynew}.
Unfortunately, all of these models
have unwanted fractionally charged states that could
survive in their light spectra.
This is to be expected, since there is a general result
\cite{Schellekens} that if a string model is to completely avoid
fractionally-charged color-neutral string states,
then its affine levels must obey the relation
\beq
 3\,k_Y + 3\,k_2 + 4 \,k_3 ~=~ 0~~~ ({\rm mod}~ 12)~.
\label{schellconstraint}
\eeq
For $k_2=k_3 =1,2$, this implies $k_Y/k_2\geq 5/3$.
Of course, it is possible that fractionally charged states
appear but are extremely massive, or that they might bind together into
color-neutral objects under the influence of extra hidden-sector
interactions.  A general classification of the binding scenarios that
can eliminate such fractionally charged states has been
performed~\cite{k1paper}, but no string model has yet been
constructed which realizes these scenarios.

 {\it Heavy string threshold corrections:}~~
Heavy string threshold corrections are the contributions due to the
infinite towers of massive Planck-scale string states that are
otherwise neglected when deriving a low-energy effective
action from the string.
In order to reconcile the values of the three
low-energy gauge couplings $g_i$ with string-scale
unification, it turns out that such corrections $\Delta_i$ must
have the relative sizes
\beq
          \Delta_{\hat Y}-\Delta_2 \approx -28    ~,~~~
          \Delta_{\hat Y}-\Delta_3 \approx -58    ~,~~~
          \Delta_{2}-\Delta_3 \approx -30    ~.
\label{finaldiffsreqd}
\eeq
where $\hat Y\equiv Y/\sqrt{5/3}$ is the renormalized hypercharge.
These corrections are quite sizable, and the fundamental question
is then how to obtain corrections of this size.

The formalism for calculating these corrections
was first derived in Ref.~\cite{Kaplunovsky} and more recently refined in
Ref.~\cite{Kiritsis}.  From these results,
a number of theoretical mechanisms were identified for making these
corrections sufficiently large.
Perhaps the most obvious mechanism~\cite{DKL}
is to construct a string model with a large modulus (such as a large
compactified dimension), for
as the size of such a radius is increased,
various momentum states become lighter and lighter.
The contributions of such states to the threshold corrections
therefore become more substantial, ultimately leading
to a decompactification of the theory.
Unfortunately, it is not known why a given string model
should be expected to have such a large modulus.
Indeed, the general expectation is that
in realistic string models,
moduli should settle at or near the self-dual
point for which moduli are of order one~\cite{nolargemoduli}.

Explicit calculations of these threshold corrections
have been carried out within several realistic string models.
Here the term ``realistic'' denotes string models with
the following properties:
$N=1$ spacetime SUSY;  appropriate gauge groups [such
as $SU(3)\times SU(2)\times U(1)$, Pati-Salam $SO(6)\times SO(4)$,
or flipped $SU(5)\times U(1)$];  the proper massless observable
spectrum (including three complete chiral MSSM generations
with correct quantum numbers, hypercharges, and Higgs scalar
representations);  and anomaly cancellation.
Many realistic models also exhibit additional attractive
features, such as a semi-stable proton, proper fermion mass hierarchy,
and a heavy top quark.
A collection of such models, all of which
are realized in the free-fermionic construction
with an underlying $\IZ_2\times \IZ_2$ orbifold structure,
can be found in Ref.~\cite{models}.

Unfortunately, the results found within these models are not
encouraging:
in each of the realistic string models of Ref.~\cite{models},
it is found~\cite{unif1}
that threshold corrections are unexpectedly small,
and moreover they have the wrong sign.
For example, in one such string model it was found
that
\beq
\Delta_{\hat Y}- \Delta_2 ~\approx~ 1.6~,~~~~
          \Delta_{\hat Y}- \Delta_3 ~\approx~ 5~
\eeq
which does not fare well against Eq.~(\ref{finaldiffsreqd}).
This behavior seems to be generic to the entire class
of realistic string models in Ref.~\cite{models}.
Thus it seems that threshold corrections by themselves
are not able to resolve the discrepancy with the
low-energy couplings in
these realistic string models.
Indeed, despite some interesting proposals~\cite{NS}, there do not
presently exist any realistic string models with small moduli
for which the threshold corrections are sufficiently large.

 {\it Light SUSY thresholds and intermediate-scale
       gauge structure:}~~
Light SUSY thresholds are the effects that arise from SUSY-breaking
at some intermediate scale:  they can be parametrized in terms
of the usual soft SUSY-breaking parameters $\lbrace m_0, m_{1/2},
m_h, m_{\tilde h}\rbrace$, or one can
take non-universal boundary terms for the sparticle masses.
Similarly, the effects from intermediate-scale gauge structure
arise whenever there is a gauge symmetry, such as $SO(6)\times SO(4)$
or flipped $SU(5)\times U(1)$, which is broken at some intermediate
scale $M_I$.  Such effects are then parametrized in terms of $M_I$.
Both of these effects are analyzed purely in terms of the
low-energy field theory derived from the string, and
consequently their evaluation proceeds exactly as in field theory.
A detailed calculation of these effects
must also include two-loop corrections,
the effects of Yukawa couplings, and even the effects of
scheme conversion (from the $\overline{\rm DR}$ scheme in
which the string scale is evaluated to the $\overline{\rm MS}$
scheme through which the low-energy couplings are extracted
from experiment).
Within the context of the low-energy effective theories derived
from the realistic string models in Ref.~\cite{models},
such a calculation has been performed~\cite{unif1}.
The results indicate that the light SUSY
thresholds are generally insufficient to resolve the discrepancies,
and that the effects of intermediate gauge structure in the
realistic string models only {\it enlarge}\/ the disagreement
with experiment!
This latter result is surprising, given that $M_I$ can be tuned
in principle to {\it any}\/ value below $M_{\rm string}$, and serves
to illustrate the rather tight (and predictive) constraints that
a given string model provides.

  {\it Extra Matter Beyond the MSSM:}~~
Finally, there is the possibility of extra matter beyond the MSSM.
While all of the above results assumed only the MSSM spectrum,
string theory often {\it requires}\/ that additional exotic states
appear in the massless spectrum.
Their effects must therefore be included.  Such states appear
in a majority of the realistic string models,  usually appear
in vector-like representations, and   ultimately have masses
determined by cubic and higher-order terms in the superpotential
(which are determined in turn by the specific SUSY-breaking mechanism
employed, as well as by a host of additional factors).
In one string model, for example, it has been estimated~\cite{massestimate}
that such extra states
will naturally sit at an intermediate scale $\approx 10^{11}$ GeV.
In the realistic string models~\cite{models} with $SU(3)\times
SU(2)\times U(1)$ gauge groups,
such matter typically arises in rather specific $SU(3)\times SU(2)\times
U(1)_Y$
representations
such as $ ({\bf 3},{\bf 2})_{1/6}$,
   $(\overline{{\bf 3}},{\bf 1})_{1/3}$,
   $(\overline{{\bf 3}},{\bf 1})_{1/6}$,
   and $({\bf 1},{\bf 2})_{0}$.
While the first two representations can be fit into standard $SO(10)$
mutliplets, the remaining two cannot, and are truly exotic.

What is remarkable, however, is that this extra matter is just what
is needed:  because of their unusual hypercharge assignments, these
representations have one-loop beta-function coefficients $b_i$ where
$b_1$ turns out to be much smaller than $b_2$ or $b_3$.
These representations therefore have the potential to modify the
running of the $SU(2)$  and $SU(3)$ couplings without seriously affecting
the $U(1)$ coupling.  Moreover, in some string models, these extra
non-MSSM matter representations also appear in precisely the
right {\it combinations}\/ to do the job.
Details can be found in Ref.~\cite{unif1}.
Similar scenarios using such extra non-MSSM matter can also
be found, {\it e.g.}, in Ref.~\cite{othermatter}.
Thus, on the basis of this evidence,
it appears that extra intermediate-scale matter beyond the MSSM
may turn out to be the string-preferred
route to string-scale unification.
It is remarkable that string theory, which predicts an unexpectedly
high unification scale, often also simultaneously predicts precisely
the extra exotic matter necessary to reconcile this higher scale
with the observed low-energy couplings.

\separator

\setcounter{footnote}{0}

 ~

\noindent
{\large\bf Postscript}

 ~

In this article, we have surveyed a number of questions
and issues that arise in supersymmetric particle physics,
ranging from the MSSM at the lowest scales to string theory at
the highest scales.
It is remarkable that supersymmetry not only provides
a window into physics at so many different energy regimes,
but also has such a profound impact in all of these areas.
Indeed, at the very least it
either refines old questions or proposes new ones,
and in most cases it actually changes the language of the debate.
Supersymmetry
is perhaps the only extension of the Standard Model which
has such a direct impact on so many types of new phenomena,
including gravity.  Moreover, as we have seen, supersymmetry has
global applications, ranging from high-energy accelerator
experiments to astrophysics and cosmology.
The questions that supersymmetry prompts
 therefore provide unique opportunities
for studying all sorts of new physics,
and finding the answers to any of these questions --- from the most
phenomenological to the most theoretical --- will undoubtedly
teach us much about the physics that we expect to be
exploring in the twenty-first century.

\separator


 ~

\noindent
{\large\bf Acknowledgments}

 ~

We would like to thank
J.~Bagger,
A.~Kusenko,
P.~Langacker,
J.~March-Russell, and
N.~Polonsky
for many helpful discussions, and especially
K.S.~Babu,
T.~Gherghetta,
G.~Kane, and
M.~Peskin for their comments on this article.
This work was supported in part by DOE Grant No.\ DE-FG02-90ER40542.
CK would also like to acknowledge the generous support of Helen and
Martin Chooljian, and the hospitality of the
Aspen Center for Physics where parts of this work were completed.

\vfill\eject


\end{document}